\documentclass[PAPER, atlasdraft=false, USenglish, cernpreprint, texmf, orcidlogo]{atlasdoc}
\usepackage{atlaspackage}
\usepackage{atlasbiblatex}

\usepackage{comment}

\usepackage[hepprocess=true]{atlasphysics}
\usepackage{atlasparticle}
\usepackage{atlasjetetmiss}

\usepackage{makecell}
\usepackage{multirow}

\graphicspath{{logos/}{figures/}}

\usepackage{ANA-HDBS-2021-07-PAPER-defs}
\usepackage{multirow}
\usepackage[export]{adjustbox}
\AtlasTitle{Search for decays of the Higgs boson into a pair of pseudoscalar particles decaying into $b\bar{b}\tau^+\tau^-$ using $pp$ collisions at $\sqrt{s}=$ 13 TeV with the ATLAS detector}

\AtlasAbstract{%
This paper presents a search for exotic decays of the Higgs boson into a pair of new pseudoscalar particles, $H\rightarrow aa$, where one pseudoscalar decays into a $b$-quark pair and the other decays into a $\tau$-lepton pair, in the mass range $12\leq m_{a}\leq 60$ GeV. The analysis uses $pp$ collision data at $\sqrt{s} = 13$ TeV collected with the ATLAS detector at the LHC, corresponding to an integrated luminosity of $140~\mathrm{fb}^{-1}$. No significant excess above the Standard Model (SM) prediction is observed. Assuming the SM Higgs boson production cross-section, the search sets upper limits at 95\% confidence level on the branching ratio of Higgs bosons decaying into $b\bar{b}\tau^+\tau^-$, $\mathcal{B}$($H \rightarrow aa \rightarrow b\bar{b}\tau^+\tau^-$), between 2.2\% and 3.9\% depending on the pseudoscalar mass.
}

\AtlasRefCode{HDBS-2021-07}

\PreprintIdNumber{CERN-EP-2024-164}

\arXivId{2407.01335}

\AtlasJournalRef{Phys.\ Rev.\ D 110 (2024) 052013}
\AtlasDOI{10.1103/PhysRevD.110.052013}

\HepDataRecord{152754}

\AtlasJournal{Phys.\ Rev.\ D.}
\hypersetup{pdftitle={ATLAS document},pdfauthor={The ATLAS Collaboration}}

\begin{document}

\maketitle

\section{Introduction}

Following the observation of the Higgs boson $H$ with mass $m_H$ near $125~\GeV$ by the ATLAS and CMS collaborations~\cite{HIGG-2012-27,CMS-HIG-12-028}, studies of its properties have been important programs of research.  Global analyses of measurements of Higgs boson properties constrain the branching ratio of the Higgs boson into undetected Beyond the Standard Model (BSM) particles to approximately $\mathcal{B}_\text{u} \lesssim 12\%$~\cite{HIGG-2021-23,CMS-HIG-22-001}. Higgs boson decays are particularly sensitive to new physics due to the small total width ($\Gamma_H \approx 4~\MeV$). Even very small couplings to new particles can give sizable branching ratios and can be compatible with available measurements~\cite{Curtin:2013fra}.

\begin{figure*}[!bp]
\begin{center}
\includegraphics[width=0.35\textwidth,valign=c]{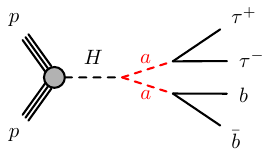}
\caption{Feynman diagram for the leading contribution to the $pp \rightarrow H \rightarrow aa \rightarrow b\bar{b}\tau^+\tau^-$ process.}
\label{fig:feyn_diags}
\end{center}
\end{figure*}

Extensions of the Standard Model (SM) that include new light pseudoscalars, called $a$-bosons, can give rise to exotic Higgs boson decays $H \rightarrow aa$. Such new light particles appear in theories with an extended Higgs sector~\cite{Dobrescu:2000yn,Ellwanger:2003jt,Dermisek:2005ar,Chang:2008cw,Morrissey:2008gm},
dark matter models~\cite{Silveira:1985rk,Pospelov:2007mp,Draper:2010ew,Ipek:2014gua,Martin:2014sxa},
models with a first-order electroweak phase transition~\cite{Kozaczuk:2019pet,Carena:2019une},
and theories of neutral naturalness~\cite{Craig:2015pha,Curtin:2015fna}.
Signatures of $H \rightarrow aa$ can also arise in models with hidden-sector particles that are singlets under the SM gauge transformations~\cite{Curtin:2013fra,Robens:2015gla,Robens:2016xkb,Robens:2019kga,Bauer:2017ris}. In scenarios where the $a$-boson mixes with the SM Higgs boson and inherits its Yukawa couplings to fermions, decays of the $a$-boson into heavy fermions such as $b$-quarks and $\tau$-leptons are favored, and the process $H \rightarrow aa \rightarrow b\bar{b}\tau^+\tau^-$, shown in Figure~\ref{fig:feyn_diags}, is expected to have a sizeable branching ratio in the mass range  $2 m_{b} < m_{a} < m_H/2$~\cite{Curtin:2013fra,Haisch:2018kqx}.

This paper presents a search for the exotic Higgs boson decay \haabbtautau and uses the full Run-2 dataset of $pp$ collisions at $\sqrt{s}=13~\TeV$ recorded with the ATLAS detector corresponding to an integrated luminosity 140 \ifb. The analysis is performed over the mass range $12 < m_{a} < 60~\GeV$ and targets the following production modes of the Higgs boson: gluon-gluon fusion (ggF), vector boson fusion (VBF), and associated production with a vector boson ($VH$). The associated production of a top-antitop-quark pair ($t\bar{t}$) with a Higgs boson provides a negligible signal contribution due to (i) the low cross-section and (ii) the additional $b$-quarks from the top-quark decays resulting in a reduced signal acceptance. Therefore, this production mode is not considered. Different analysis categories are defined depending on the $\tau$-lepton decay modes, which can be into electrons ($e$), muons ($\mu$), or hadrons ($\tauh$, also ``hadronic taus''). Due to the relatively low mass of the Higgs boson and the four-body final state, the $\tau$-leptons and $b$-quarks tend to have low transverse momentum ($\pt$), typically below $\pt \lesssim 50\,\GeV$.   Therefore, the analysis relies on electron or muon triggers, which have lower \pt thresholds compared to triggers using hadronic activity, to select the events of interest and requires an electron or muon in the final state. The analysis defines different event categories depending on $\tau$-lepton decay modes: $e\mu$, $e\tauh$, and $\mu\tauh$.
The major backgrounds to the signal process are Drell--Yan production of $\tau$-leptons produced in association with heavy-flavor jets, $t\bar{t}$, and non-prompt leptons plus $\tauh$.

For low $m_a$, the $a$-boson has a large Lorentz boost and its decay products can be collimated. Consequently, the $a\rightarrow b\bar{b}$ final state is reconstructed as a single jet that contains the hadronization products of the two $b$-quarks. While several techniques exist to resolve merged jets~\cite{Larkoski:2017jix}, most of them are only efficient in the case of high-mass and high-$\pt$ jets. Due to the relatively low mass of the Higgs boson, the merged $a\to b\bar{b}$ jet will have low $\pt$. This analysis uses a novel, dedicated algorithm to identify low-mass, merged, ``double $b$-quark'' jets ($B$-jets)~\cite{ATL-PHYS-PUB-2022-042}. For high $m_a$, the $b$-quarks decays tend to be well separated and the reconstructed jets capture the hadronization of a single $b$-quark ($b$-jet).
The analysis considers events with one $B$-jet, or one or two $b$-jets, resulting in nine analysis categories, as shown in Figure~\ref{fig:categories}.

\begin{figure*}[!tbp]
\begin{center}
\includegraphics[width=0.4\textwidth,valign=c]{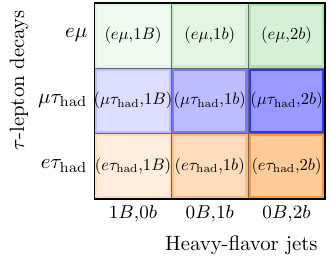}
\caption{The analysis categories used in the search corresponding to different decay modes of the $\tau$-leptons and strategies for identifying heavy-flavor jets, including merged \bjet pairs ($B$) and single $b$-jets ($b$). Note that the categories are exclusive.}
\label{fig:categories}
\end{center}
\end{figure*}

Similar searches in the $b\bar{b}\tau^+\tau^-$ decay channel were performed  by the CMS Collaboration~\cite{CMS-HIG-17-024, CMS-HIG-22-007}. The latest search has placed 95\% CL upper limits on \braabbtautau in the range $1.7-7.7$\% for $12\leq m_{a}\leq 60\,\GeV$ using 138 \ifb\ of Run 2 data at $\sqrt{s} = 13$~\TeV. This analysis improves the sensitivity of previous results \cite{CMS-HIG-22-007} in the low mass regime ($m_a<20$~\GeV) by targeting more final states and using a neural network discriminant to increase the separation of signal from background. This search is also complementary to other searches for $H \rightarrow aa$ decays performed by the ATLAS and CMS collaborations using both $\sqrt{s} = 8~\TeV$ and $\sqrt{s} = 13~\TeV$ data in several final states including $\mu^+\mu^-\mu^+\mu^-$~\cite{EXOT-2013-15,EXOT-2016-22,HDBS-2018-55},
$\mu^+\mu^-\tau^+\tau^-$~\cite{HIGG-2014-02,CMS-HIG-17-029,CMS-HIG-18-006,CMS-HIG-18-024},
$\tau^+\tau^-\tau^+\tau^-$~\cite{CMS-HIG-14-019,CMS-HIG-18-006},
$b\bar{b}\mu^+\mu^-$~\cite{HIGG-2016-03,HDBS-2021-03,CMS-HIG-16-015,CMS-HIG-18-011},
$b\bar{b}b\bar{b}$~\cite{HIGG-2016-01,HIGG-2017-05,HDBS-2018-47},
$\gamma\gamma\gamma\gamma$~\cite{EXOT-2013-24,CMS-HIG-21-003,CMS-HIG-21-016},
and $\gamma\gamma gg$~\cite{HIGG-2017-09}.

\section{ATLAS detector}

The ATLAS detector~\cite{PERF-2007-01} at the LHC covers nearly the entire solid angle around the collision point.\footnote{ATLAS uses a right-handed coordinate system with its origin at the nominal interaction point (IP)
in the centre of the detector and the \(z\)-axis along the beam pipe.
The \(x\)-axis points from the IP to the centre of the LHC ring,
and the \(y\)-axis points upwards.
Polar coordinates \((r,\phi)\) are used in the transverse plane,
\(\phi\) being the azimuthal angle around the \(z\)-axis.
The pseudorapidity is defined in terms of the polar angle \(\theta\) as \(\eta = -\ln \tan(\theta/2)\) and is equal to the rapidity
$ y = \frac{1}{2} \ln \left( \frac{E + p_z c}{E - p_z c} \right) $ in the relativistic limit.
Angular distance is measured in units of \(\Delta R \equiv \sqrt{(\Delta y)^{2} + (\Delta\phi)^{2}}\).}
It consists of an inner tracking detector surrounded by a thin superconducting solenoid, electromagnetic and hadronic calorimeters,
and a muon spectrometer incorporating three large superconducting air-core toroidal magnets.

The inner-detector system (ID) is immersed in a \qty{2}{\tesla} axial magnetic field
and provides charged-particle tracking in the range \(|\eta| < 2.5\).
The high-granularity silicon pixel detector covers the vertex region and typically provides four measurements per track,
the first hit generally being in the insertable B-layer (IBL) installed before Run~2~\cite{ATLAS-TDR-19,PIX-2018-001}.
It is followed by the SemiConductor Tracker (SCT), which usually provides eight measurements per track.
These silicon detectors are complemented by the transition radiation tracker (TRT),
which enables radially extended track reconstruction up to \(|\eta| = 2.0\).
The TRT also provides electron identification information
based on the fraction of hits (typically 30 in total) above a higher energy-deposit threshold corresponding to transition radiation.

The calorimeter system covers the pseudorapidity range \(|\eta| < 4.9\).
Within the region \(|\eta|< 3.2\), electromagnetic calorimetry is provided by barrel and
endcap high-granularity lead/liquid-argon (LAr) calorimeters,
with an additional thin LAr presampler covering \(|\eta| < 1.8\)
to correct for energy loss in material upstream of the calorimeters.
Hadronic calorimetry is provided by the steel/scintillator-tile calorimeter,
segmented into three barrel structures within \(|\eta| < 1.7\), and two copper/LAr hadronic endcap calorimeters.
The solid angle coverage is completed with forward copper/LAr and tungsten/LAr calorimeter modules
optimised for electromagnetic and hadronic energy measurements respectively.

The muon spectrometer (MS) comprises separate trigger and
high-precision tracking chambers measuring the deflection of muons in a magnetic field generated by the superconducting air-core toroidal magnets.
The field integral of the toroids ranges between \num{2.0} and \qty{6.0}{\tesla\metre}
across most of the detector.
Three layers of precision chambers, each consisting of layers of monitored drift tubes, cover the region \(|\eta| < 2.7\),
complemented by cathode-strip chambers in the forward region, where the background is highest.
The muon trigger system covers the range \(|\eta| < 2.4\) with resistive-plate chambers in the barrel, and thin-gap chambers in the endcap regions.

The luminosity is measured mainly by the LUCID--2~\cite{LUCID2} detector that records Cherenkov light produced in the quartz windows of photomultipliers located close to the beampipe.
Events are selected by the first-level trigger system implemented in custom hardware,
followed by selections made by algorithms implemented in software in the high-level trigger~\cite{TRIG-2016-01}.
The first-level trigger accepts events from the \qty{40}{\MHz} bunch crossings at a rate below \qty{100}{\kHz},
which the high-level trigger further reduces in order to record complete events to disk at about \qty{1}{\kHz}.
A software suite~\cite{SOFT-2022-02} is used in data simulation, in the reconstruction
and analysis of real and simulated data, in detector operations, and in the trigger and data acquisition
systems of the experiment.

\section{Data and simulated event samples}

This search uses $pp$ collision data at $\sqrt{s} = 13$ \TeV\ recorded by the ATLAS experiment from 2015 to 2018.
Only events that satisfy data quality requirements that ensure the stable operation of the ATLAS detector~\cite{DAPR-2018-01} are considered. The resulting dataset corresponds to an integrated luminosity of $140.1 \pm 1.2~\ifb$~\cite{DAPR-2021-01}. Data are selected using a combination of single-electron, and single-muon, and opposite-flavor dilepton ($e\mu$) triggers~\cite{TRIG-2018-05, TRIG-2018-01}.

Several Monte Carlo (MC) event generators are used to simulate the signal and background processes. The MC samples are used to optimize the event selection, evaluate efficiencies and acceptances, and to estimate yields. The main SM backgrounds are Drell--Yan production with decays into $\tau$-leptons $Z/\gamma^{*} \rightarrow \tau^+\tau^-$ produced in association with jets ($Z+$jets), top-quark production ($\ttbar$ or single top quarks) where at least one of the $W$ bosons decay leptonically, and backgrounds where jets are misidentified as leptons. Other backgrounds from Higgs bosons decaying into $\tau$-leptons $H \rightarrow \tau^+ \tau^-$, $Z/\gamma^{*} \rightarrow \ell^+\ell^-~(\ell=e, \mu)$, diboson production, and vector bosons produced in association with $\ttbar$ are also included.

Samples of $Z+$jets and $W+$jets events were produced using the \SHERPA[2.2.11]~\cite{Bothmann:2019yzt} generator with next-to-leading-order (NLO) precision matrix elements for up to two partons, and leading-order (LO) precision for up to five partons in the five-flavor scheme calculated with the \textsc{Comix}~\cite{Gleisberg:2008fv} and \OPENLOOPS~\cite{Buccioni:2019sur,Cascioli:2011va,Denner:2016kdg} libraries.
The events were matched with the \SHERPA parton shower~\cite{Schumann:2007mg} using the \MEPSatNLO prescription~\cite{Hoeche:2011fd,Hoeche:2012yf,Catani:2001cc,Hoeche:2009rj} with a dedicated set of tuned parameters.
The \NNPDF[3.0nnlo] set of parton distribution functions (PDF)~\cite{Ball:2014uwa} was used in the sample generation and the samples were normalized to a next-to-next-to-leading-order (NNLO) prediction~\cite{Anastasiou:2003ds}.

Samples of diboson (\(VV\)) events were produced with the \SHERPA[2.2.1] generator for semileptonic final states or \SHERPA[2.2.2]~\cite{Bothmann:2019yzt} generator for fully leptonic final states.
Fully leptonic final states and semileptonic final states, where one boson decays leptonically and the other hadronically, were generated using matrix elements at NLO accuracy in QCD for up to one additional parton and at LO accuracy for up to three additional parton emissions.
Samples for the loop-induced processes \(gg \to VV\) were generated using LO-accurate matrix elements for up to one additional parton emission for both the cases of fully leptonic and semileptonic final states.
The matrix element calculations were matched and merged with the \SHERPA parton shower based on Catani--Seymour dipole factorization~\cite{Gleisberg:2008fv,Schumann:2007mg} using the MEPS@NLO prescription. The virtual QCD corrections were provided by the \OPENLOOPS library~\cite{Buccioni:2019sur,Cascioli:2011va,Denner:2016kdg}.
The \NNPDF[3.0nnlo] PDF set was used~\cite{Ball:2014uwa}, along with a dedicated set of tuned parton-shower parameters.

The production of \ttbar events was modeled using the \POWHEGBOX[v2]~\cite{Frixione:2007nw,Nason:2004rx,Frixione:2007vw, Alioli:2009je} generator at NLO precision in QCD with the \NNPDF[3.0nlo] PDF set and the \hdamp\footnote{The \hdamp parameter regulates singularities in the emission of hard radiation in \POWHEG. The value is chosen to provide good description of the \ttbar system \pt.} parameter was set to 1.5\,$m_{t}$~\cite{ATL-PHYS-PUB-2016-020}, with the top-quark mass $m_{t}$ set to 172.5~\GeV.
The parton shower, hadronization, and underlying event were modeled using \PYTHIA[8.230]~\cite{Sjostrand:2014zea} with the A14 set of tuned parameters~\cite{ATL-PHYS-PUB-2014-021} and using the \NNPDF[2.3lo] PDF set~\cite{Ball:2012cx}. The decays of bottom and charm hadrons were performed by \EVTGEN[1.6.0]~\cite{EvtGen}.

Single-top-quark production events were modeled with dedicated samples covering $s$-channel, $t$-channel, or $\Wboson$-associated ($tW$) production. All three production modes were modeled using \POWHEGBOX[v2]~\cite{Nason:2004rx,Frixione:2007vw,Alioli:2010xd,Alioli:2009je,Frederix:2012dh,Re:2010bp} at NLO in QCD with the \NNPDF[3.0nlo] PDF set. For the $s$-channel and $tW$ production, the calculation is performed in the five-flavor scheme, while the $t$-channel production uses the four-flavor scheme. Parton shower and hadronization of these events were modeled with \PYTHIA[8.230] using the A14 tune and the \NNPDF[2.3lo] PDF set.

The signal event samples include Higgs boson production via ggF, VBF, and $VH$. For the SM production of the $H\rightarrow \tau^+\tau^-$ background process, only the ggF and VBF production modes are considered.  Top-quark-associated production modes of the SM $H\rightarrow \tau^+\tau^-$ background process are negligible.

The sample of Higgs boson production via ggF is generated at NLO accuracy in QCD using \POWHEGBOX[v2]~\cite{Hamilton:2013fea,Hamilton:2015nsa,Alioli:2010xd,Nason:2004rx,Frixione:2007vw}.
The simulation achieves NNLO accuracy for $gg\to h$ observables by reweighting the Higgs boson
rapidity spectrum in \textsc{Hj}-\MINLO~\cite{Hamilton:2012np,Campbell:2012am,Hamilton:2012rf} to that of HNNLO~\cite{Catani:2007vq}.
The Higgs boson production via VBF and $VH$ were simulated with NLO precision using \POWHEGBOX[v2]~\cite{Nason:2009ai,Alioli:2010xd,Nason:2004rx,Frixione:2007vw}. The \PDFforLHC[15nnlo] PDF set~\cite{Butterworth:2015oua} was used for all three production modes, while the parton shower and hadronization were performed using \PYTHIA[8.244] with the A14 set of tuned parameters.

Each of the simulated Higgs production samples are normalized using dedicated higher-order cross-section calculations.
The ggF samples are normalized to the next-to-next-to-next-to-leading-order (N$^3$LO) cross-section in QCD plus electroweak corrections at next-to-leading logarithm (NLL)~\cite{deFlorian:2016spz,Anastasiou:2016cez,Anastasiou:2015ema,Dulat:2018rbf,Harlander:2009mq,Harlander:2009bw,Harlander:2009my,Pak:2009dg,Actis:2008ug,Actis:2008ts,Bonetti:2018ukf,Bonetti:2018ukf}.
The VBF samples are normalized to an approximate-NNLO QCD cross-section with NLO electroweak corrections~\cite{Ciccolini:2007jr,Ciccolini:2007ec,Bolzoni:2010xr}.
Similarly, the $VH$ samples are normalized to cross-sections calculated at NNLO in QCD with NLO electroweak corrections. The cross section estimate includes the $gg\to ZH$ contribution, even though it is not simulated explicitly.
For the signal samples, the decay \haabbtautau is performed using \PYTHIA[8.244] and eight distinct mass points were simulated for the ggF, VBF, and the $VH$ production modes in the range $m_a = 12 - 60$~GeV (12, 16, 20, 25, 30, 40, 50 and 60~\GeV).

All simulated event samples were processed through a detailed simulation of the ATLAS detector based on $\GEANT$~\cite{GEANT4:2002zbu} or a faster simulation where the full \GEANT simulation of the calorimeter response is replaced by a detailed parameterization of the shower shapes~\cite{SOFT-2010-01}.
The effects of multiple $pp$ interactions in the same and neighboring bunch crossings (\pileup) were modeled by overlaying the simulated hard-scattering event with inelastic $pp$ (minimum-bias) events to reproduce the pileup distributions seen in the data.
These inelastic events were generated with the soft QCD processes of \PYTHIA[8.186]~\cite{Sjostrand:2007gs} using the \NNPDF[2.3lo] PDF set and the A3 set of tuned parameters~\cite{ATL-PHYS-PUB-2016-017}.

\section{Object and event reconstruction}
\label{sec:objectReconstruction}

Events are required to contain at least one reconstructed $pp$ collision vertex candidate with at least two associated ID tracks with $\pT > 0.5~\GeV$~\cite{ATL-PHYS-PUB-2015-026}. The primary vertex (PV) is selected as the vertex with the highest sum of the squared transverse momentum of the associated tracks.

Electrons are reconstructed from a seed cluster in the EM calorimeter matched to a track in the ID~\cite{EGAM-2018-01} and are required to have $\pt>7~\GeV$ and $|\eta|<2.47$. Electrons in the calorimeter barrel--endcap transition region ($1.37 <|\eta| < 1.52$) are excluded from the analysis. Electrons from $\tau$-lepton decays can be displaced from the PV due to the long $\tau$-lepton lifetime, so no requirement on the transverse impact parameter ($d_0$) is applied. A requirement on the longitudinal impact parameter $|z_0\sin\theta|<0.5$~mm is applied to match the electron track to the PV of the event. Electrons must satisfy the \textit{Medium} working point of the likelihood identification criteria~\cite{EGAM-2018-01}. Electrons satisfying these requirements are referred to as baseline electrons. Signal electrons are defined by additionally applying a tight isolation requirement using both tracking and calorimetry information with $\pT$-dependent $\Delta R$ cone radius ~\cite{EGAM-2018-01}. This isolation variable rejects electrons that likely originated from light- or heavy-flavor hadrons. The isolation criterion is corrected for the presence of a nearby muon by subtracting the $\pt$ of the muon track within the isolation cone of the electron from the isolation sum. This is particularly important for low mass signal samples ($m_a\lesssim20~\GeV$), where the electrons and muons from two $\tau$-lepton decays in $a\rightarrow \tau^+\tau^-$ are often found to be within the isolation cone of each other.

Several methods are used to reconstruct muons, depending on the availability of tracks in the ID and the MS \cite{MUON-2018-03}. In order to benefit from the extended MS coverage up to $|\eta|=2.7$, an ID track is not required in the region $2.5<|\eta|<2.7$. Muons are selected with $\pt > 7~\GeV$ and $|\eta|<2.7$. As in the case of electrons, the only impact parameter requirement applied is $|z_0\sin\theta|<0.5$~mm. Muons must satisfy the \textit{Medium} identification criterion~\cite{MUON-2018-03}, comprising the baseline muon selection. Signal muons are defined with an additional loose requirement on the isolation in both the inner tracking detector and calorimeters also with a $\pt$-dependent cone radius~\cite{MUON-2018-03}. As above, an electron within the isolation cone of a muon is excluded from the calculation.

Jets are reconstructed using the \antikt algorithm \cite{Cacciari:2008gp} implemented in the \Fastjet package~\cite{Fastjet,Cacciari:2005hq} with a radius parameter $R=0.4$. A particle-flow approach~\cite{PERF-2015-09} is used for the jet reconstruction. The jet energy scale is calibrated to the particle level using simulation and further corrected with in-situ methods~\cite{JETM-2018-05}. The jet selection requires $\pT > 15~\GeV$ and $|\eta|<2.5$. A multivariate jet vertex tagger (JVT) based on tracking information is used to identify a jet as originating from the PV and suppress jets from pileup interactions for jets with $\pt<60~\GeV$ and $|\eta|<2.4$~\cite{PERF-2014-03}.

The strategy to identify jets containing $b$-hadrons depends on the kinematics of the signal. For low signal masses ($m_a\lesssim20~\GeV$), pairs of $b$-hadrons from $a\to b\bar{b}$ decays tend to be merged in the detector and are identified as a single reconstructed jet. \DEXTER~\cite{ATL-PHYS-PUB-2022-042} is an end-to-end algorithm that classifies jets into three categories: merged $b$-jets, single $b$-jets, and other jet flavors. Track-jets associated to jets are reconstructed by reclustering jets with $\pt > 20~\GeV$ and $|\eta| < 2.0$, together with all ID tracks using an \antikt algorithm with radius parameter $R=0.8$. When the track-jet contains a single $R=0.4$ jet, the jet is called isolated. An exclusive-$k_t$ clustering of the track-jet tracks into precisely two sub-jets (ex-$k_t^{(2)}$)~\cite{Ellis:1993tq} is used to reconstruct the flight axes of the two $b$-jets within a track-jet. Both ex-$k_t^{(2)}$ track-subjets are required to have $\pt > 5~\GeV$, where the transverse momentum of the subjet is estimated by summing the four-momentum of the associated tracks. The \DEXTER algorithm uses the tracks from the $R=0.8$ track-jets to reconstruct secondary vertices, which provide a distinctive signature of merged $b$-jets when more than one is reconstructed inside the same jet or when they merge in a single secondary vertex with very high mass. Displaced tracks, secondary vertices, and the properties of the two ex-$k_t^{(2)}$ jets are used in a deep set neural network (NN) to classify the flavor of the jet. The NN exploits the presence of highly displaced tracks along the two flight axes as well as the reconstruction of multiple secondary vertices with large mass to resolve the two $b$-jets.

In the case of jets which are not merged, the jet flavor is identified with the \dlr algorithm~\cite{FTAG-2019-07} combining track impact parameter values with information from secondary vertices reconstructed within the jet. A working point corresponding to $60\%$ efficiency for identifying $B$-jets is used for \DEXTER, and a working point corresponding to $85\%$ efficiency for identifying $b$-jets is used for \dlr.

The \bjet identification efficiency of the \dlr\ algorithm is measured in collider data by using $\ttbar$, $Z$+jets, $W$+jets, and multijet events~\cite{FTAG-2018-01, FTAG-2020-08, FTAG-2019-02}. A dedicated measurement of the \dlr algorithm identification efficiency for jets with $15<\pt<20~\GeV$ was performed for this search. The $B$-jet identification efficiency of the \DEXTER\ algorithm is also measured in collider data by using \ttbar\ and $Z$+$g(\to \bbbar)$ events~\cite{ATL-PHYS-PUB-2022-042}. These measurements are used to correct the identification efficiency of heavy-flavored jets in simulation. The calibration procedure corrects the identification efficiency in simulation as a function of the jet $\pt$ and $\eta$ to match the one observed in data. Discrepancies between the identification efficiency in simulation and in data stem from both mismodeling of the detector response and of the underlying physics processes.

The reconstruction of $\tauh$ candidates is seeded using jets reconstructed using the \antikt algorithm with distance parameter $R=0.4$ whose inputs are topoclusters, three-dimensional clusters of calorimeter cells~\cite{PERF-2014-07}.
Reconstructed $\tauh$ candidates have $\pt>20~\GeV$ and $|\eta|<2.5$, excluding the region $1.37 <|\eta| < 1.52$. The identification of $\tauh$ uses a recurrent neural network (RNN) algorithm, which uses as inputs tracks and calorimeter clusters associated to $\tauh$ candidates, as well as high-level discriminating variables~\cite{ATL-PHYS-PUB-2019-033}. Baseline $\tauh$ candidates are required to pass the \textit{Very Loose} working point, while signal $\tauh$ candidates are further required to satisfy the \textit{Medium} working point~\cite{ATL-PHYS-PUB-2019-033}.
A dedicated multivariate electron veto is applied to $\tauh$ candidates to reject electrons misidentified as $\tauh$.
The \textit{Medium} working point is used for the boosted-decision-tree-based electron veto~\cite{ATL-PHYS-PUB-2015-045}.

A dedicated $\tau$-jet-vertex algorithm (TJVA) is used to associate $\tauh$ candidates to a PV~\cite{PERF-2013-06}. The algorithm does not apply impact parameter requirements and finds the vertex with the largest fraction of the $\pt$ from the tracks associated with the $\tauh$ within a distance of $R=0.2$ around the candidate.
While TJVA does not apply impact parameter requirements, tracks must be sufficiently close to the $\tauh$ vertex, so two selections are applied on the impact parameters of tracks: $|d_0^{\text{TJVA}}|<1.0$~mm and $|z_0^{\text{TJVA}}\sin\theta|<1.5$~mm~\cite{ATLAS-CONF-2017-029}.

An overlap removal procedure is applied to prevent double counting of objects. This procedure is applied to the objects described previously that satisfy the baseline criteria, with the exception of muons where candidates satisfying the \textit{VeryLoose} identification criterion are used for the overlap removal and the \textit{Medium} identification criteria is used after overlap removal for the baseline selection criteria. This reduces backgrounds from $Z/\gamma^{*} \rightarrow \mu^+\mu^-$ events where a \textit{VeryLoose} muon can be misidentified as a $\tauh$ candidate.

The overlap removal procedure is executed as follows. The closest jet within a radius of $\Delta R=0.2$ of a selected electron is removed. If the nearest jet is within $\Delta R = 0.4$ of an electron, the electron is excluded from the analysis. Muons are excluded from the analysis if they are separated from the nearest jet by $\Delta R < 0.4$, since this reduces the background from heavy-flavor decays inside jets. However, if the jet has fewer than three associated tracks, the muon is selected and the jet is excluded from the analysis instead. This avoids an inefficiency for high-energy muons undergoing significant energy loss in the calorimeter. Electrons are excluded from the analysis if they share their track with a muon.

From the remaining jets, isolated jets that satisfy the 60\% \DEXTER working point are classified as $B$-jets. Isolated jets that do not satisfy the 60\% \DEXTER working point and satisfy the 85\% \dlr working point are classified as selected $b$-jets. Isolated jets that do not satisfy the 60\% \DEXTER working point and do not satisfy the 85\% \dlr working point are not used in the analysis. Non-isolated jets that satisfy the 85\% \dlr working point are also classified as $b$-jets, while those that fail are excluded from the analysis. Muons in the annular region $0.4 < \Delta R < 0.8$ region around a $B$-jet which have a relative calorimeter-based isolation larger than 0.1 are excluded. Electrons satisfying the same criteria are also removed from the analysis. Finally, any $B$-jet with either an electron or a muon in the annular $0.4 < \Delta R < 0.8$ region and relative calorimeter-based isolation less than 0.1 are excluded from the analysis.

Muons that are excluded in the overlap removal procedure, but that are within $\Delta R < 0.3$ of a $b$-jet or within $\Delta R < 0.3$ of either of the ex-$k_t^{(2)}$ track-subjets of a $B$-jet are classified as soft muons. The leading soft muon associated to a $b$-jet or to a ex-$k_t^{(2)}$ track-subjet is added to the jet four momentum to account for semileptonic decays of $b$-hadrons in the jet energy.

The missing transverse momentum \vecmet, with magnitude \met, is calculated\ using the magnitude of the vector sum of the $\pt$ of all reconstructed objects and any additional tracks not associated to any reconstructed objects in the event~\cite{JETM-2020-03}. The missing transverse energy (\met) measures the imbalance of the transverse momentum in the detector. In this search \met is expected to arise from the neutrinos from $\tau$-lepton decays.

\section{Event selection}
\label{sec:eventSelection}

Events are placed in exclusive categories defined by the final state objects from the $\tau^+\tau^-$ and $b\bar{b}$ systems, as shown in Figure~\ref{fig:categories}. The trigger strategy used to select events depends on the leptons in the final state. For the \emu categories, three orthogonal trigger regions are defined. If the event has at least one electron with $\pt>27~\GeV$, the single electron trigger is used and is required to match the selected electron. If the event has no such electron, but has a muon with $\pt>27~\GeV$, the single muon trigger is equivalently used and is required to match the selected muon. If there are no electrons nor muons satisfying the previous criteria, then opposite-flavor $e\mu$ triggers are used, also requiring trigger matching for both leptons. For the $e\tauh$ and $\mu\tauh$ categories, the single electron or muon trigger of the corresponding flavor with $\pt>27~\GeV$ is used and matched to the electron or muon respectively.

Events in the \emu categories are required to have exactly one electron and one muon with opposite-sign (OS) charge that satisfy the signal selection criteria. In addition, events should have at most one baseline and no signal $\tauh$ candidate. In the $e\tauh$ and $\mu\tauh$ categories, events are required to have exactly one electron and one muon, respectively, and exactly one signal $\tauh$ with OS charge. For the \emu categories, the electron and muon are required to be separated by $\DeltaR(e,\mu)>0.1$ and for the $e\tauh$ and $\mu\tauh$ categories, this requirement is $\DeltaR(e/\mu, \tauh)>0.2$, to reduce backgrounds from low-mass hadronic decays. The analysis categories are also defined by requiring one $B$-jet, or one or two $b$-jets.

Two additional criteria are applied in each category to define regions enhanced in backgrounds. These criteria are based on the kinematics of the visible $\tau$-lepton decay products, i.e., excluding the neutrinos. The visible mass of the $\tau$-lepton decay products $m^{\text{vis}}(\tau\tau)$ is lower for signal than for the $Z\rightarrow\tau^+\tau^-$ background, due to the lower mass of the $a$-boson.  A requirement is thus applied on the visible mass for all categories: $m^{\text{vis}}(\tau\tau) = m^{\text{vis}}(e\mu) < 45~\GeV$ for the \emu categories and $m^{\text{vis}}(\tau\tau) = m^{\text{vis}}(e\tauh)$ or $m^{\text{vis}}(\mu\tauh)  < 60~\GeV$ for the $e\tauh$ and $\mu\tauh$ categories. The difference in the requirements is due to the smaller number of neutrinos in the categories with a hadronic $\tau$-lepton compared to the \emu case.  A low mass requirement on the visible mass of $m^{\text{vis}}(\tau\tau)>4~\GeV$ is also applied to reduce backgrounds from hadronic decays. The sample obtained by inverting the higher requirement on the visible mass is used to correct the modeling of the $Z$+jets background (see Section~\ref{sec:backgroundEstimation}), and is called the ``$Z$ region.''
Figure~\ref{fig:SR_variables}(a) demonstrates the ability of the background prediction to model the data in the most sensitive $\muhad$ category.
The background model performs similarly well in the $\ehad$ and $\emu$ categories.

\begin{figure}[!tbp]
\centering
\subfloat[]{
\includegraphics[width=0.45\textwidth]{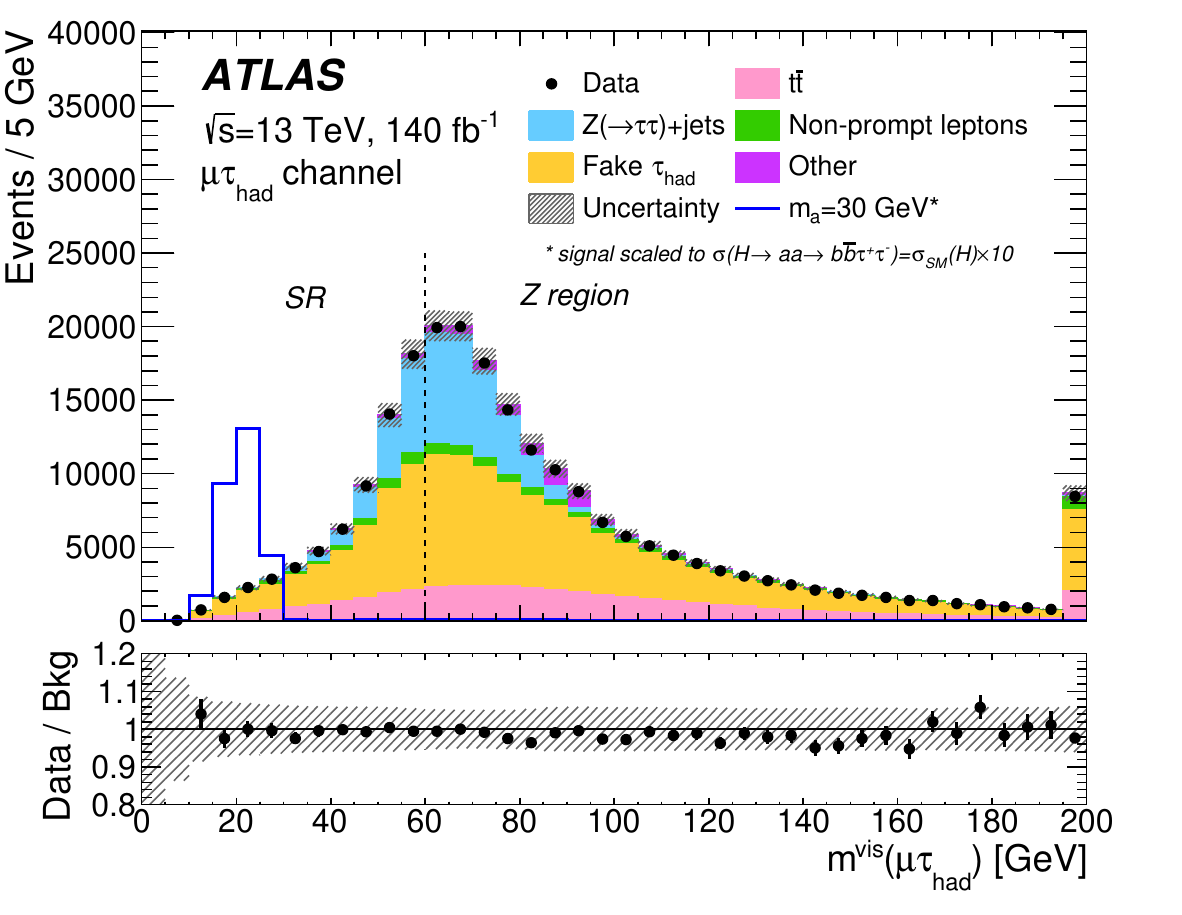}}
\subfloat[]{
\includegraphics[width=0.45\textwidth]{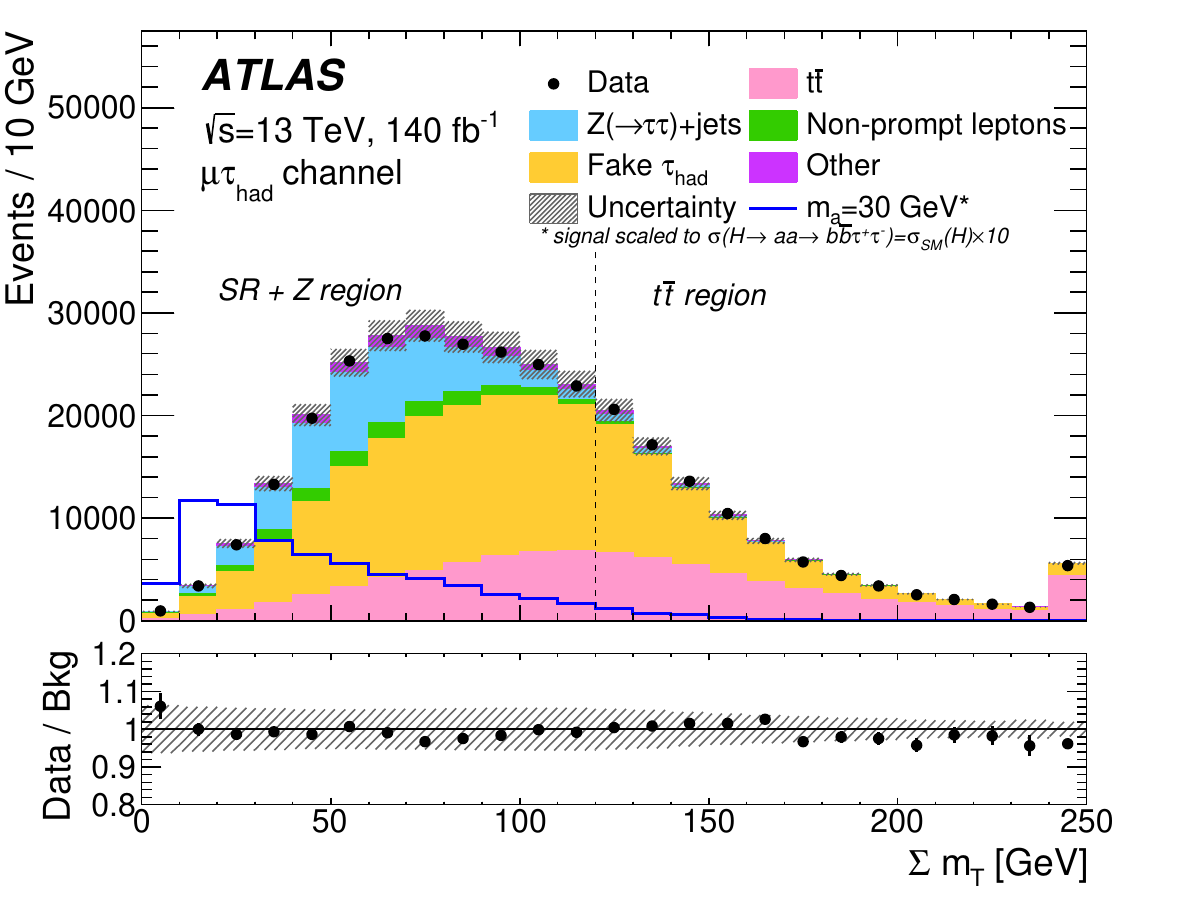}}
\caption{(a) Visible mass $m^{\text{vis}}(\mu\tauh)$ and (b) sum of the transverse mass $\Sigma m_T$ distributions for signal and the expected background. Events with high $m^{\text{vis}}(\mu\tauh)$ and high $\Sigma m_T$ are included in the $\ttbar$ region. In order to compare the shapes, the expected signal distribution is shown assuming ten times the production cross section of the Higgs boson and a 100\% branching ratio to $b\bar{b}\tau^+\tau^-$. The yields for the backgrounds correspond to the values obtained after applying the corrections to the background modeling described in Section~\ref{sec:backgroundEstimation}. The cuts separating the signal region (SR) from the $Z$ and \ttbar regions are indicated by the vertical dashed line. The hashed area represents the total uncertainty of the background. Overflow events are included in the last bins.}
\label{fig:SR_variables}
\end{figure}

When compared to the background, leptons from signal processes also have lower transverse mass calculated with the $\tau$-lepton visible transverse momentum $p_{\text{T}}^{\text{vis}}(\tau)$ and $E_T^{\text{miss}}$. In the case of leptonic decays of the $\tau$-lepton, the visible $\pt$ is defined as the transverse momentum of the electron or muon. The transverse mass is defined as: $ m_{\mathrm{T}}(\tau) = \sqrt{2 \pt^{\text{vis}}(\tau) \met (1-\cos{(\Delta\phi)})}$, where $\Delta\phi$ is the difference in the azimuthal angle between $p_{T}^{\text{vis}}(\tau)$ and $E_T^{\text{miss}}$. A requirement is applied on the sum of the transverse mass calculated for the two $\tau$-leptons $\Sigma \mT<120~\GeV$, where $\Sigma \mT=\mT(\tau^{\text{lead}}) + \mT(\tau^{\text{sublead}})$. Figure~\ref{fig:SR_variables}(b) shows the distribution of $\Sigma\mT$ in the $\mu\tauh$ categories. The sample obtained by inverting the requirement on $\Sigma \mT$ and is used to correct the modeling of the $\ttbar$ background (see Section~\ref{sec:backgroundEstimation}), and is called the ``$\ttbar$ region.''  There is small signal contamination into the $Z$ and $t\bar{t}$ regions, only $\approx 1$ signal event is expected assuming $\mathcal{B}(\haabbtautau)$=10\%.

Finally a same-sign (SS) region, enhanced in backgrounds containing jets that are misidentified as electrons, muons or hadronic taus, is defined by applying the same selection criteria as the signal region but requiring the $e$, $\mu$. and $\tauh$ to have same-sign charge. This region is used to estimate backgrounds with non-prompt leptons (see Section~\ref{sec:backgroundEstimation}). Table~\ref{tab:selectionCuts} summarizes the event selection used in the analysis.

\begin{table}
\caption{Event selection for the analysis categories and the background regions. The background regions only list the requirements that are different from the signal region. The definition of signal leptons can be found in Section~\ref{sec:objectReconstruction}.}%
\label{tab:selectionCuts}
\centering
\begin{tabular}{lcc} \toprule
Region & \emu & $e\tauh$ or $\mu\tauh$ \\
\midrule
\multirow{7}{*}{Signal region}    & 1 OS signal \emu pair & 1 OS signal $e\tauh$ or $\mu\tauh$ pair \\
& 0 signal $\tauh$     & 1 signal $\tauh$             \\
& $\Delta R(e,\mu)>0.1$ & $\Delta R(\ell, \tau)>0.2$\\
& $4<m^{\text{vis}}(\tau\tau)<45~\GeV$ & $4<m^{\text{vis}}(\tau\tau)<60~\GeV$\\
& \multicolumn{2}{c}{$\Sigma m_T<120$~\GeV}\\
& \multicolumn{2}{c}{1 $B$-jet or 1 or 2 $b$-jets} \\
\midrule
$Z$ region & $m^{\text{vis}}(\tau\tau)>45~\GeV$ & $m^{\text{vis}}(\tau\tau)>60~\GeV$ \\
\midrule
\ttbar region & \multicolumn{2}{c}{$\Sigma m_T>120~\GeV$, no $m^{\text{vis}}(\tau\tau)$ requirement}\\
\midrule
SS region & 1 SS signal \emu pair & 1 SS signal $e\tauh$ or $\mu\tauh$ pair\\
\bottomrule
\end{tabular}
\end{table}

\section{Background estimation}
\label{sec:backgroundEstimation}

Several SM processes can produce final states that satisfy the object and event selections described previously. The contributions from these processes are estimated with simulation or data-driven methods. Background processes with all leptons originating from the decay of vector bosons are defined as prompt and described by simulation. The dominant sources of prompt backgrounds are top-quark production, from \ttbar and single-top events, and $Z+$jets where the $Z$ boson decays into pairs of $\tau$-leptons. Other backgrounds estimated from simulation include $h\to\tau^+\tau^-$, diboson production, $\ttbar$ in association with a vector boson, and $Z/\gamma^* \rightarrow \ell^+\ell^-$ where one lepton is misidentified as a $\tauh$ in the $e\tauh$ and $\mu\tauh$ categories.

The modeling of two of the main prompt backgrounds, $Z+$jets and $\ttbar$, in simulation is corrected to match the data in regions enhanced in these backgrounds (see Section~\ref{sec:eventSelection}). For the \ttbar background, the simulation is reweighted to match the number of $b$-jets and the number of jets in the $\ttbar$ region. The \ttbar region is devoid of $Z+$jets events and performing the $\ttbar$ correction first minimizes the need for further iterations. In the \emu categories, an additional correction is applied as a function of the variable $H_T$, defined as the scalar sum of the $\pt$ of all jets and leptons in the event. The reweighting procedure improves the description of the $\pt$ of the $\ttbar$ system and the description of the number of additional jets produced in association with the $\ttbar$ system. The same strategy was applied in other ATLAS searches with large $\ttbar$ background~\cite{HDBS-2018-51,HDBS-2020-16}. After the $\ttbar$ reweighting, the $Z+$jets background is corrected with two normalization factors. The $Z\to \tau^+\tau^-$ and $Z\to e^+e^-$ backgrounds are scaled so that the predicted yields from simulation in two $m_{\ell\ell}$ bins match the observed number of events in the $Z$ region.

Background processes with at least one lepton originating from the non-prompt decay of a hadron, from photon conversions, or from the misidentification of other particles are defined as non-prompt backgrounds. A tight-to-loose data-driven method~\cite{ATLAS:2022swp} is used to estimate this background contribution. As described in Section~\ref{sec:objectReconstruction}, each lepton has two criteria baseline and signal. Leptons that satisfy the baseline selection but not the signal one are called loose ($L$), while those satisfying the signal selection are called tight ($T$). The estimates of the non-prompt backgrounds are obtained by assigning a weight $w_L=f\varepsilon/(\varepsilon-f)$ to each loose lepton and a weight $w_T=\varepsilon(1-f)/(\varepsilon-f)$ to each tight lepton~\cite{ATLAS:2022swp}. Here, $f$ and $\varepsilon$ are, respectively, the rate with which non-prompt and prompt leptons that satisfy the baseline selection also satisfy the signal criteria. The sign of the overall event weight is adjusted depending on the number of loose leptons to avoid double-counting of background events.

The non-prompt rates for electrons and muons are measured with events in the SS region that satisfy $\Delta R(e,\mu)>1.4$. Events with one baseline muon (electron) and one signal electron (muon) are used to estimate the muon (electron) non-prompt rate $f_{\mu}$ ($f_e$). The electron and muon non-prompt rates are parameterized as a function of the lepton $\pt$, $\eta$, whether or not the lepton is matched to a trigger object, and the number of $b$-jets in the event. The prompt rates for electrons (muons) $\epsilon_e$ ($\epsilon_\mu$) are estimated as a function of the same variables from simulations of $\Zboson\to\tau^+\tau^-$ events.

The condition $\Delta R(e,\mu)>1.4$ ensures that the leptons do not interfere in their identification efficiency. When the weights are applied in the $e\mu$ signal region, a geometrical correction is necessary to account for the overlap of isolation cones with $\Delta R(e,\mu) < 0.6$. The correction is applied to events with two loose leptons and is parameterized as:

\begin{align}
(-w^{e}_Lw^{\mu}_L)^{\text{corr}}&=\left(1-f(\DeltaR)\right)\left(-w^{e}_Lw^{\mu}_L\right) + f(\DeltaR)\times\frac{1}{2}\left(w^{e}_Lw^{\mu}_T+w^{e}_Tw^{\mu}_L\right),  \nonumber \\
f(\DeltaR)&=c\times\left(\frac{2}{\pi}\arccos\left(\frac{\DeltaR}{2r}\right)-\frac{\DeltaR}{\pi r^2}\sqrt{r^2-\frac{(\DeltaR)^2}{4}}\right), \nonumber
\end{align}
where $f(\DeltaR)$ is the fraction of the isolation cone that overlaps, $c$ is an arbitrary constant fit from the data in the SS region, and $r=0.3$ is the radius of the isolation cone.

For hadronic taus, the prompt rate is taken as $\varepsilon_{\tau}=1$ in the weights $w_L$ and $w_T$, and MC simulation is used to remove the contribution from prompt loose taus. The non-prompt rate is measured separately with three different processes: $Z(\to \mu^+\mu^-) + \tauh$, $\text{jet}+\tauh$ where the $\tauh$ candidate has high JVT score, and $\text{jet}+\tauh$ with low JVT score. In each region, the non-prompt rate is parameterized as a function of the $\tauh$ $\pt$, $\eta$, decay mode, and the number of $b$-jets in the event. The $\tauh$ non-prompt rate used to estimate the non-prompt background is written as a linear combination of the three estimates:

\begin{equation}
f_{\tau} = [f_{Z\to \mu^+\mu^-}(1-r_{\text{QCD}})+f_{\text{high JVT}}r_{\text{QCD}}](1-r_{\text{LJVT}}) + f_{\text{low JVT}}r_{\text{LJVT}} \nonumber,
\end{equation}
where the coefficient $r_{\text{QCD}}$ is the fraction of non-prompt background events with two non-prompt leptons in each $e\tauh$ and $\mu\tauh$ categories and the coefficient $r_{\text{LJVT}}$ is determined by a maximum-likelihood fit to the baseline $\tauh$ seed jet width distribution. The coefficients $r_{\text{QCD}}$ and $r_{\text{LJVT}}$ are determined separately for each category and region of the analysis, as well as for each $\tau$-lepton decay mode.

\section{Analysis strategy}
\label{sec:analysis_strategy}

In each analysis category, a NN is trained using the kinematic variables of the reconstructed heavy-flavor jets and $\tau$-leptons. The NN is parameterized as a function of the $a$-boson mass in order to obtain an optimal discriminant for each different simulated mass~\cite{pNN}.
During training, the background is assigned a random value as the value of $m_a$, while for signal the simulated  mass is used.
The signal events have different correlations between physical observables and the generated mass than the background, which the network exploits.
Once the network is trained, the NN output score is calculated with the true mass replaced by the mass hypothesis under consideration for all events, be they signal, background, or data.

Due to the neutrinos in the decays of the $\tau$-leptons, the four-momentum of the two $\tau$-leptons cannot be directly reconstructed. An algorithm called missing-mass calculator (MMC)~\cite{Elagin_2011} uses a Markov-chain MC to perform a maximum likelihood estimate of the neutrinos four-momenta, which is then used to reconstruct the input variables of the parameterized NN (pNN).

Beyond the true $a$-boson mass $m^{\text{true}}(\tau\tau)$, the pNNs for all categories use the visible mass $m^{\text{vis}}(\tau\tau)$, the MMC-based mass of the neutrino pair $m^{\text{MMC}}(\nu\nu)$ in each $\tau\to e\nu_{\tau}\bar{\nu}_e$ or $\tau\to \mu\nu_{\tau}\bar{\nu}_{\mu}$ decay, the missing transverse energy $\met$, the two transverse masses calculated with the visible $\pt$ of the final-state $\tau$-leptons $m_{\text{T}}(\tau)$, the leading heavy-flavor jet $\pt(b^{\text{lead}})$, and the visible transverse momentum $\pt^{\text{vis}}(\tau\tau b^{\text{lead}})$. The variable $D_\zeta$~\cite{CMS-HIG-13-021}, defined as:

\begin{equation}
D_\zeta = [\vec{p}_T^{\text{miss}}-0.85\vec{p}_{T}^{\text{vis}}(\tau\tau)]\cdot\vec{\zeta} \nonumber,
\end{equation}
where $\vec{\zeta}$ is a unit vector in the direction of the bisector between the two $\tau$-leptons, is also used as input to the network in all categories. $D_\zeta$ provides an estimate of the misalignment between the missing transverse momentum and the visible $\tau\tau$ subsystem.

In categories with two $b$-jets or one $B$-jet, additional input variables are used: the subleading heavy-flavor jet $\pt(b^{\text{sublead}})$, the transverse momentum $\pt(bb)$ and mass $m(bb)$ of the $bb$ subsystem, as well as the visible $m^{\text{vis}}(bb\tau\tau)$ and MMC-based $m^{\text{MMC}}(bb\tau\tau)$ mass estimates of the Higgs boson.
In categories with a $B$-jet, the four-momentum of the two ex-$k_t^{(2)}$ track-subjets are used as proxies for individual heavy-flavor jets.
Table~\ref{tab:inputNN} summarizes the variables used as input for the NN. Distributions of the pNN input variables for the ($\mu\tauh$, 2b) category are shown in Figure~\ref{fig:muhad_nn_inputs_CR}.  Distributions of the pNN($m_{a}$) output variable for a value of $m_a=30$~\GeV in the ($\muhad$, 2b) category and for $m_a=12$~\GeV in the ($\emu$, 1B) category are shown in Figure~\ref{fig:muhad_nn_outputs}.

\begin{table}[!tbp]
\centering
\caption{Neural-network input variables with a summary of the final-state property it describes.}
\begin{tabular}{l l}
\toprule
Feature & Description \\
\midrule
\multirow{3}{*}[1.2em]{$m^{\text{true}}(\tau\tau)$ }   &   During training: generated $a$-boson mass for signal MC.  Background events are assigned\\
& a random value of the eight signal masses.\\
& During testing: the mass hypothesis under consideration. \\
$m^{\text{vis}}(\tau\tau)$     &   Visible mass of the $\tau\tau$ system.\\
$\pt(\tau\tau)$     &   \pt of the $\tau\tau$ system.\\
$m^{\text{MMC}}(\nu\nu)$    &   MMC-based mass of the two neutrinos in $\tau\to e\nu_{\tau}\bar{\nu}_e$ or $\tau\to e\nu_{\tau}\bar{\nu}_{\mu}$ decays. \\
\met    &   Missing transverse energy.  \\
$m_{\text{T}}(\tau)$  & Transverse mass calculated with the visible $\pt$ of the final-state $\tau$-leptons. \\
$\pt(b^{\text{lead}})$  &   Transverse momentum of the leading final-state $b$-jet.\\
$\pt^{\text{vis}}(\tau\tau b^{\text{lead}})$ &  Visible $\pt$ of the $\tau\tau b^{\text{lead}}$ system.   \\
$D_\zeta$ &   Misalignment between the $\vec{E}_T^{\text{miss}}$ vector and the $\tau\tau$ system. \\
\bottomrule
\multicolumn{2}{c}{Categories with a $B$-jet or 2 $b$-jets} \\
\toprule
$\pt(b^{\text{sublead}})$  &   Transverse momentum of the subleading final-state $b$-jet. \\
$\pt(bb)$   &   Transverse momentum of the $bb$ system. \\
$m(bb)$     &   Mass of the $bb$ system.    \\
$m^{\text{vis}}(bb\tau\tau)$    &   Visible mass of the Higgs boson system. \\
$m^{\text{MMC}}(bb\tau\tau)$    &   MMC-based mass of the Higgs boson system.\\
\bottomrule
\end{tabular}
\label{tab:inputNN}
\end{table}

\begin{figure}[!tbp]
\centering

\subfloat[]{
\includegraphics[width=0.47\linewidth]{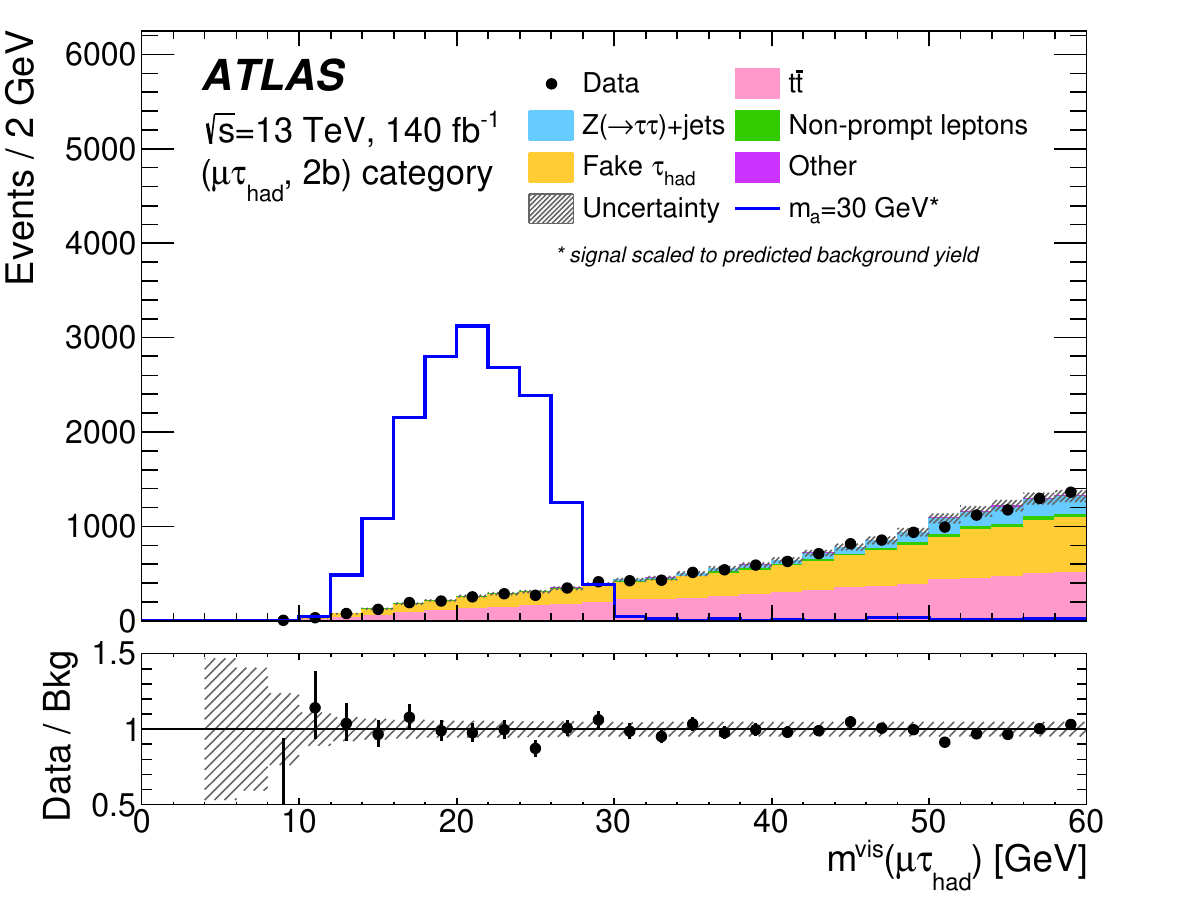}
}
\subfloat[]{
\includegraphics[width=0.47\linewidth]{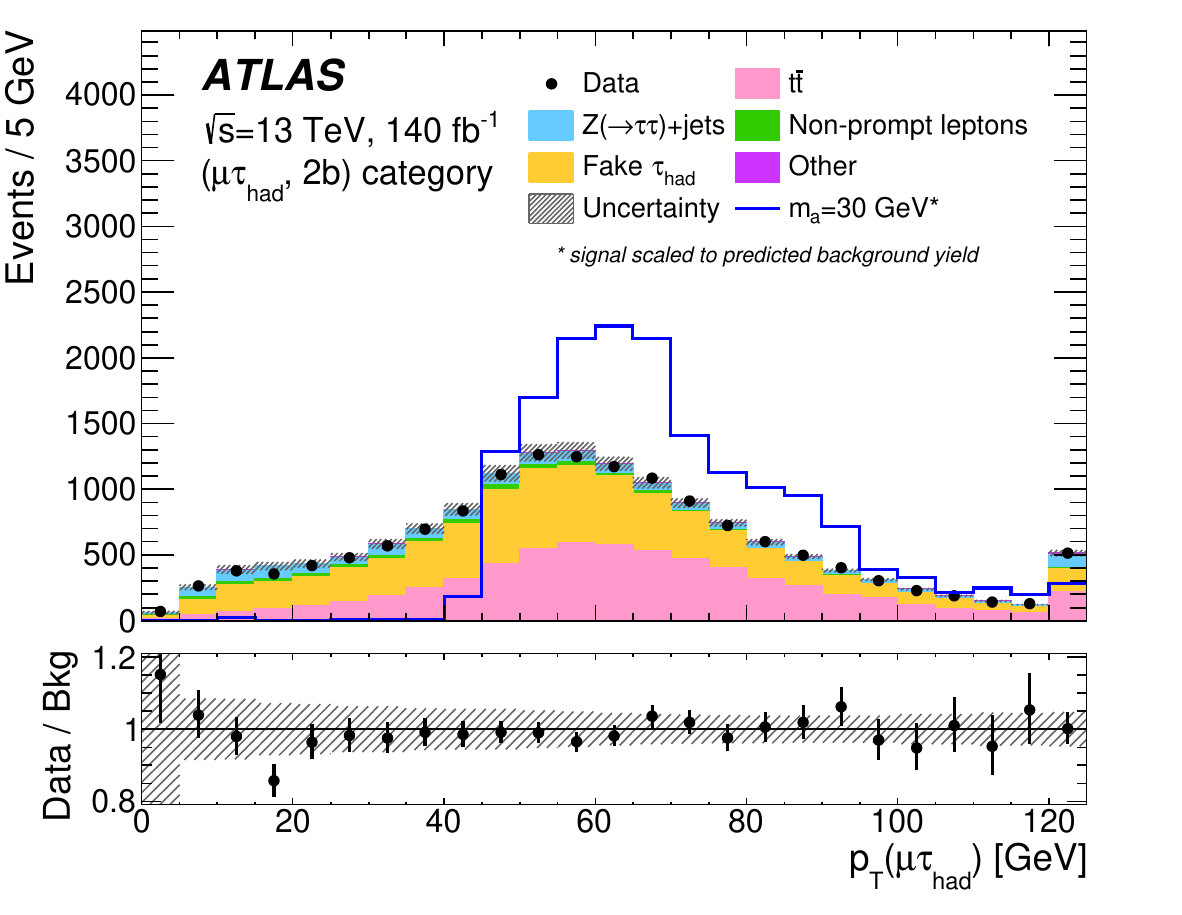}
}

\subfloat[]{
\includegraphics[width=0.47\linewidth]{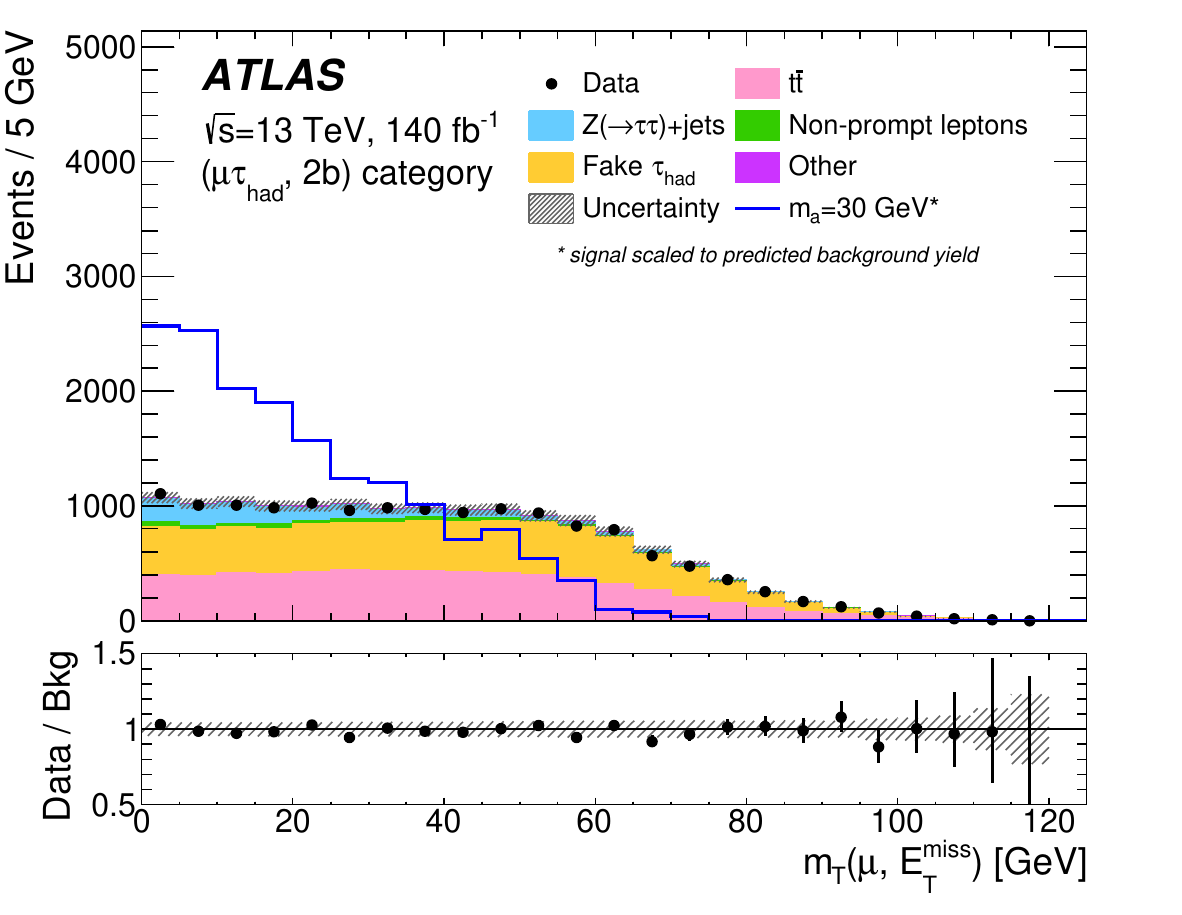}
}
\subfloat[]{
\includegraphics[width=0.47\linewidth]{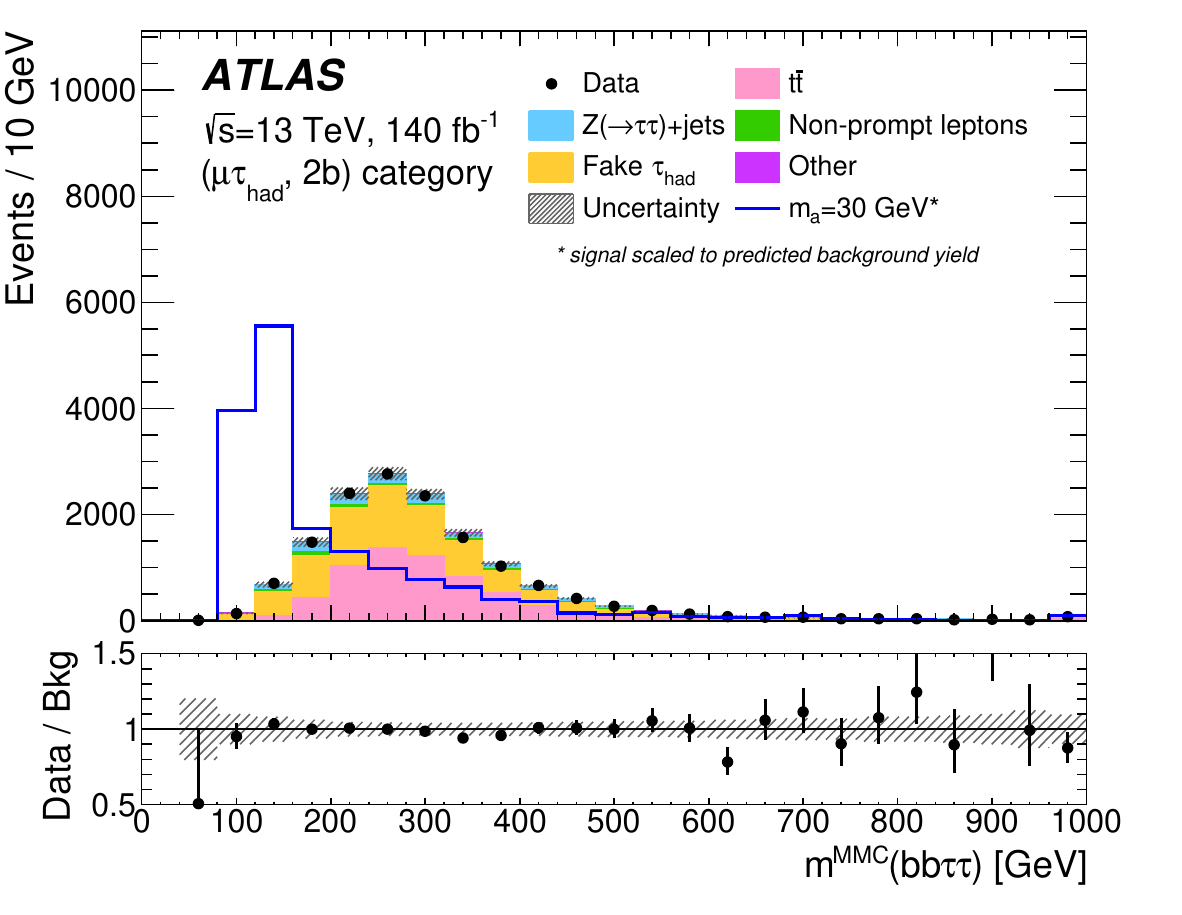}
}

\caption{The pNN input variables (a) visible mass $m^{\text{vis}}(\mu\tauh)$, (b) visible transverse momentum $\pt^{\text{vis}}(\mu\tauh)$, (c) transverse mass $m_{\text{T}}(\mu,\met)$, and (d) MMC mass $m^{\text{MMC}}(bb\tau\tau)$ are shown in the SR with no cut on the pNN discriminant. The signal shape is normalized to the same integral as the total background prediction. The hashed area represents the total uncertainty of the background.  Overflow events are included in the last bins.}%
\label{fig:muhad_nn_inputs_CR}
\end{figure}

\begin{figure}[!tbp]
\centering
\subfloat[]{
\includegraphics[width=0.47\linewidth]{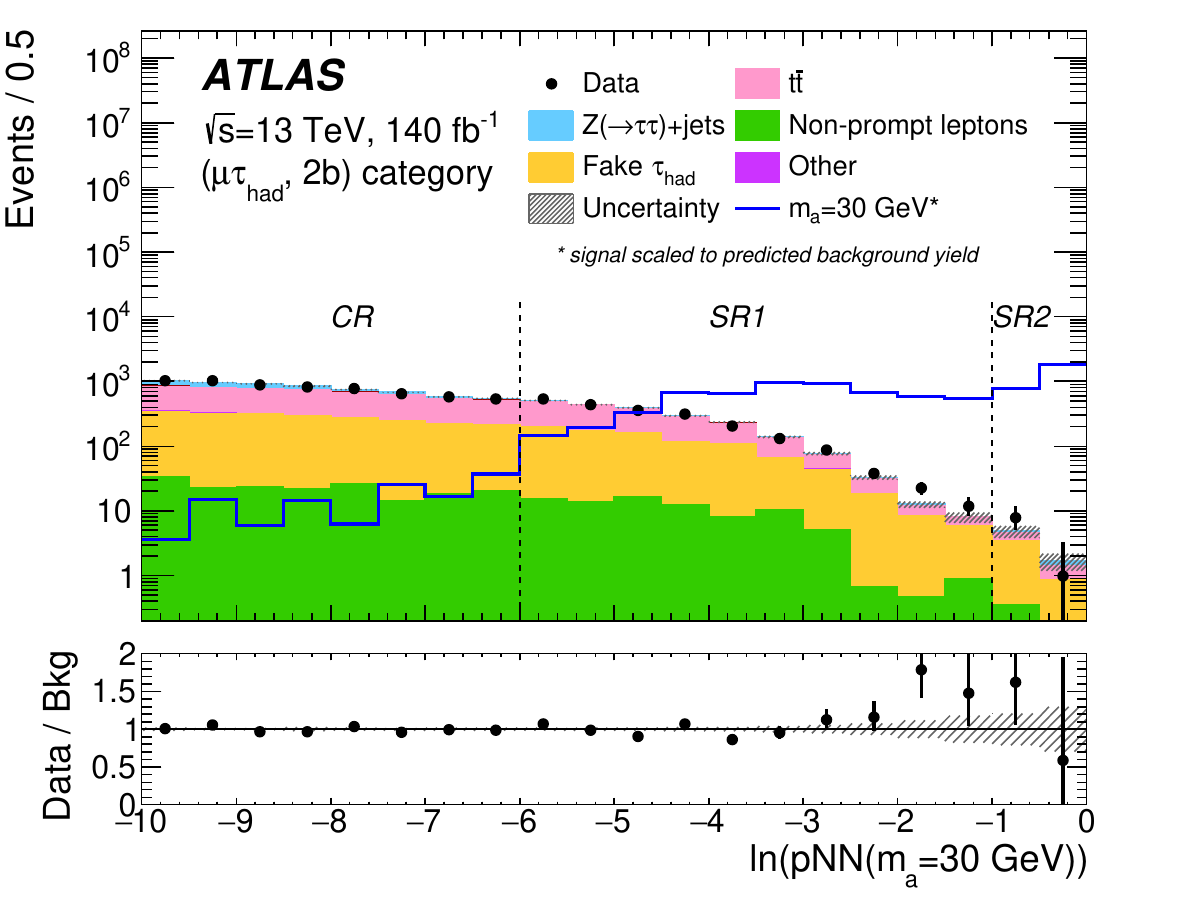}
}
\subfloat[]{
\includegraphics[width=0.47\linewidth]{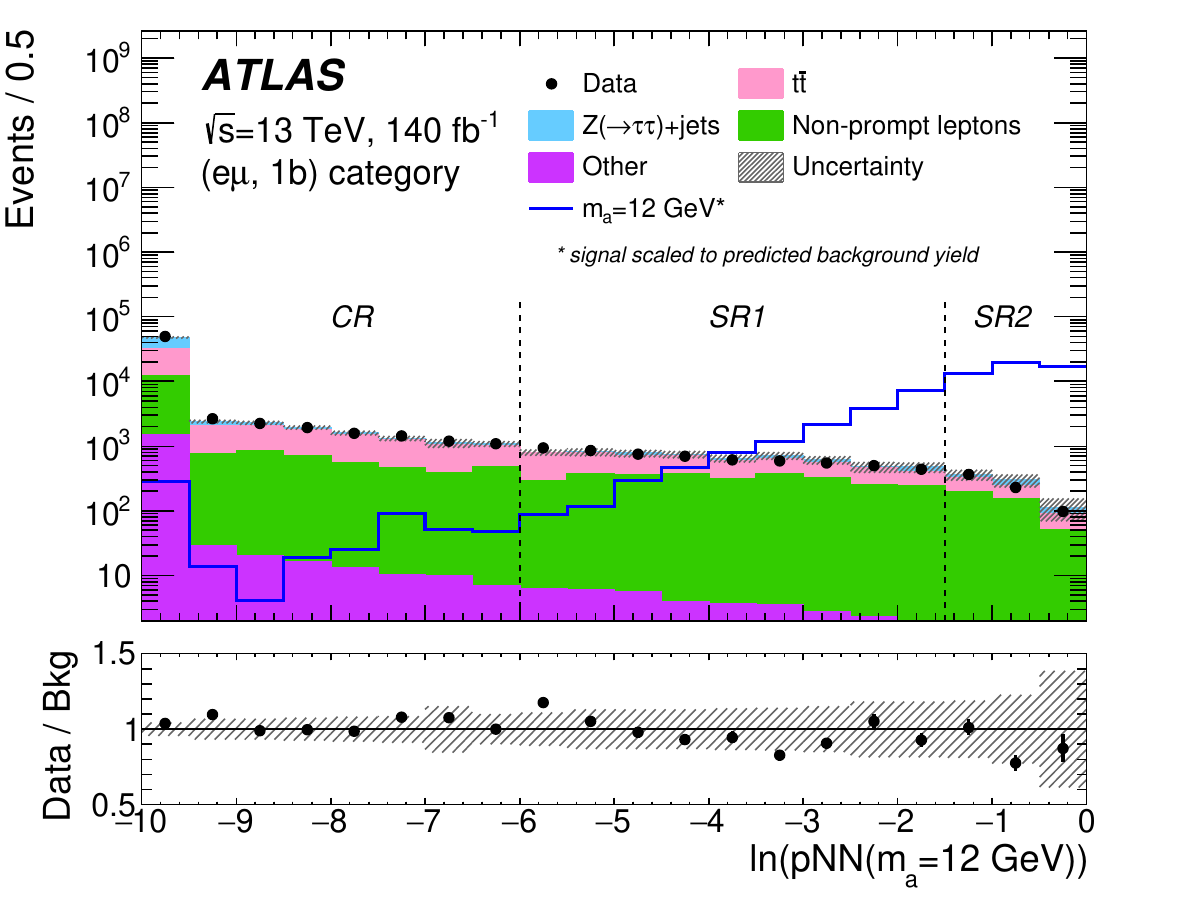}
}
\caption{Distribution of ln(pNN) output score in the (a) ($\muhad$, $2b$) category with $m_a=30$~\GeV and (b) (\emu, $1b$) category with $m_a=12$~\GeV.  The signal shape is normalized to the integral of the total background model. The hashed area represents the total uncertainty of the background. Underflow and overflow events are included in the first and last bins, respectively.}
\label{fig:muhad_nn_outputs}
\end{figure}

Each pNN is a fully connected network with three hidden layers with 15 neurons each.
The activation function is a leaky ReLU activation function with slope of \(0.01\) when the input is negative. The Objax~\cite{objax2020github} framework is used with the Adam optimization algorithm~\cite{Kingma2014AdamAM} and a binary cross entropy loss function.
The signal sample used in the NN training contains all \haabbtautau simulated samples, which are normalized so that each of them has the expected number of  observed events.
The background sample used in the NN training is composed of composed of top-quark, $Z(\rightarrow \tau\tau)+\text{jets}$ and non-prompt events.

For each $a$-boson hypothesis $m_a$, a statistical analysis is performed simultaneously in all  nine categories of the analysis. To reduce the correlation among different $m_a$ hypotheses, only events which satisfy $0.95m_a-6\,\GeV \leq m^{\text{MMC}}(\tau\tau) \leq 1.15m_a+1\,\GeV$ in the $e\tauh$ and $\mu\tauh$ categories, or $0.75m_a-6\,\GeV \leq m^{\text{MMC}}(\tau\tau) \leq 1.25m_a+1\,\GeV$ in the $e\mu$ categories are used. The pNN score calculated with this subset of events and $m^{\text{true}}(\tau\tau)=m_a$ is split into three bins with different signal-to-background ratios (S/B). These bins are called CR, SR1, and SR2, in order of increasing S/B. Figure~\ref{fig:mass_distributions} shows the ranges for each mass hypothesis tested in this paper while Figure~\ref{fig:analysis_strategy} shows the three regions used to search for a signal with $m_a=30\,\GeV$.

A test statistic is built from a profile likelihood ratio calculated using the \texttt{pyhf} software~\cite{pyhf_joss}. The likelihood is given by the product of Poisson probability distributions for the three bins in each of the nine categories and the Gaussian distributions, which implement constraint terms for each source of systematic uncertainty as a function of nuisance parameters (NP).
\begin{equation}
L(\mu,\vec{\alpha})=\prod_{c}^{N_{\text{cat}}}\prod_{j=1}^{3}\frac{(\mu s_{c,j}(\vec{\alpha}) + b_{c,j}(\vec{\alpha}))^{n_{c,j}}}{n_{c,j}!}e^{-\left(\mu s_{c,j}(\vec{\alpha}) + b_{c,j}(\vec{\alpha})\right)}\prod_{k=1}^{N_{\text{syst}}}\frac{1}{\sqrt{2\pi}}e^{-\frac{\alpha^2_k}{2}} \nonumber,
\end{equation}
where $\mu$ is the signal strength, $\vec{\alpha}$ is the vector of nuisance parameters, $s_{c,j}$ and $b_{c,j}$ are the expected number of signal and background events in the $j^{th}$ bin of the category $c$, and $n_{c,j}$ is the observed number of events. Since the signal templates are normalized to the SM Higgs inclusive cross section, the signal strength is equal to $(\sigma(H)/\sigma_{\text{SM}}(H)) \mathcal{B}(\haabbtautau)$. Note that the values $s_{c,j}$ and $b_{c,j}$ are themselves functions of the set of NPs.  Each systematic uncertainty is fully correlated across all bins and categories.

Categories with very low signal acceptance are suppressed from the likelihood product. For the $m_a=12\,\GeV$ hypothesis, only the $e\mu$ categories are used.  At this mass hypothesis, the cut $\Delta R>0.2$ between the $\tau$-leptons and lepton removes nearly all signal acceptance. For $m_a>30\,\GeV$, the categories with $B$-jets are similarly not used.  Removing these categories does not impact the observed limit due to the lack of signal acceptance.  The one-sided alternative likelihood ratio test statistic for upper limits $\tilde{q}_\mu$~\cite{Cowan:2010js} was used.

\begin{figure}
\centering
\subfloat[]{\includegraphics[width=0.5\linewidth]{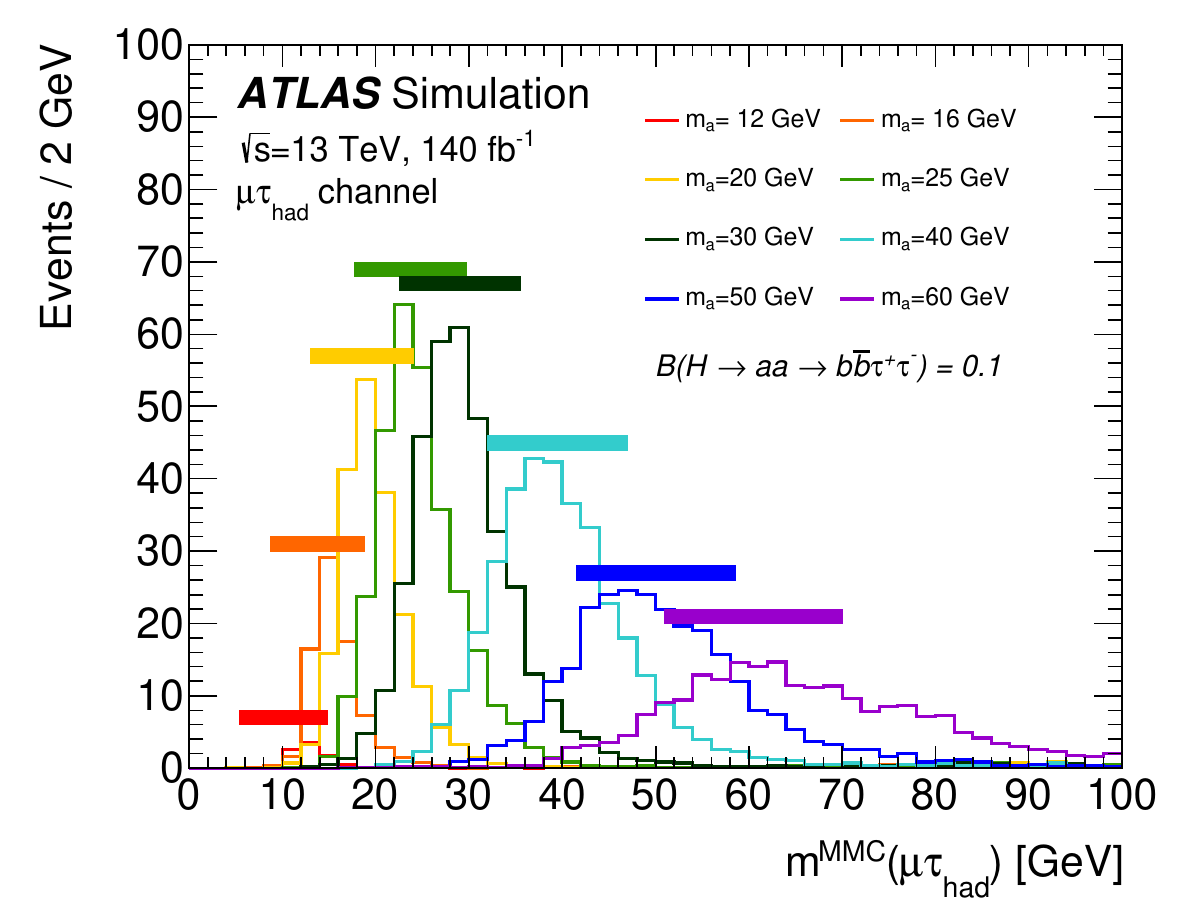}}%
\subfloat[]{\includegraphics[width=0.5\linewidth]{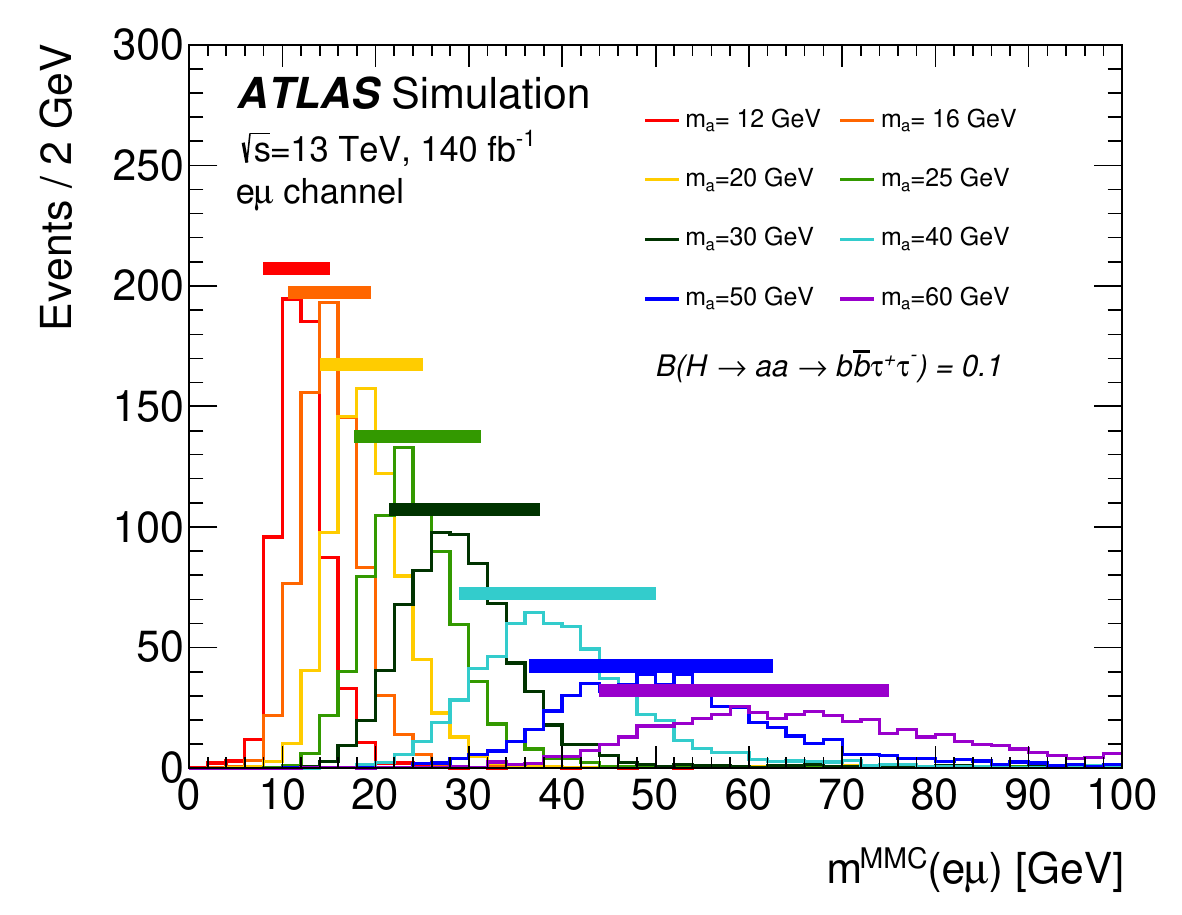}}
\caption{Mass distribution of the $\tau\tau$ system for each of the generated MC signal mass points for the (a) $\mu\tauh$ and (b) $e\mu$ categories.  The rectangle above each histogram shows the domain used to test each mass hypothesis.  Note that nearby mass hypotheses have overlapping regions. All distributions are normalized assuming $\mathcal{B}(\haabbtautau)$=10\%.}
\label{fig:mass_distributions}
\end{figure}

\begin{figure}
\centering
\includegraphics[height=6cm]{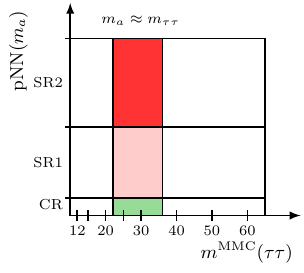}
\caption{As an illustration, the binning used in the two-dimensional plane of pNN score and mass for the $\mu\tauh$ categories and the $m_a=30$~\GeV mass hypothesis is shown.
}
\label{fig:analysis_strategy}
\end{figure}

\section{Systematic uncertainties}

Systematic uncertainties include experimental uncertainties in the reconstructed objects, uncertainties in the data-driven model for the non-prompt-lepton background, and modeling uncertainties in the simulated background and signal samples.

Experimental uncertainties include momentum scale and resolution uncertainties for all reconstructed objects: muons, electrons, hadronic $\tau$-leptons, jets, and missing transverse energy. These uncertainties are estimated by using calibrations performed for each individual object and by comparing them among different simulation models. These uncertainties are considered in the \met\ reconstruction, and additional uncertainties for soft tracks are estimated from data~\cite{JETM-2020-03}.

Uncertainties in the reconstruction, identification and trigger efficiency of electrons~\cite{EGAM-2018-01,TRIG-2018-05} and muons~\cite{MUON-2018-03,TRIG-2018-01} are determined from tag-and-probe efficiency measurements using $Z\to \ell^+\ell^-$ and $J/\psi\to \ell^+\ell^-$ events. Similarly, uncertainties in the reconstruction and identification efficiency of $\tauh$ are determined from tag-and-probe efficiency measurements using $Z\to \tau_{\mu}\tauh$ events (where $\tau_{\mu}$ denotes a $\tau \rightarrow \mu\nu_{\mu}\nu_{\tau}$ decay) and by varying the physical models used for the simulation of hadronic interactions in the detector~\cite{ATLAS-CONF-2017-029,ATL-PHYS-PUB-2019-033}. An additional uncertainty is estimated for the electron-veto efficiency based on a tag-and-probe measurement using $Z\to ee$ events~\cite{ATL-PHYS-PUB-2015-045}.

Uncertainties in the jet vertex tagger efficiency are estimated from efficiency measurements using $Z(\to\mu^+\mu^-)$ events with a single jet recoiling against the di-muon pair~\cite{PERF-2014-03}. A dedicated efficiency measurement for jets with $15\leq \pt \leq 20\,\GeV$ was performed for this search.

Uncertainties in the \dlr identification efficiency are estimated from a combination of several measurements sensitive to the true jet flavor. The efficiency is measured in several bins of jet $\pt$, $\eta$, and separately for true $b$-, $c$-, and light-jets~\cite{ FTAG-2018-01,FTAG-2020-08,FTAG-2019-02}. A dedicated efficiency measurement for jets with $15\leq \pt \leq 20\,\GeV$ was performed for this search. The measurement uses $\ttbar$ and $Z+\text{jets}$ events to measure the efficiency of the \dlr algorithm in different $\eta$ bins for $b$- and non-$b$-jets (ie., $c$- and light-jets, which are considered together in this dedicated measurement). The uncertainties in the flavor-tagging efficiency of at low jet $\pt$ are considered uncorrelated with the ones for jet $\pt > 20\,\GeV$, but the correlations between different jet $\eta$ bins and between the different true jet flavors are accounted for.

Uncertainties in the \DEXTER identification efficiency are estimated from the efficiency measurement using $\ttbar$ and $Z(\rightarrow \ell\ell) + g(\rightarrow bb)$ events~\cite{ATL-PHYS-PUB-2022-042}. The efficiency and associated uncertainties are determined in bins of jet $\pt$, $\eta$, and separately for $b$- and $B$-jets. The number of light-jets satisfying the \DEXTER identification selection  is negligible.

Modeling uncertainties for \ttbar and $Z+$jets background processes and for the Higgs boson signal process are estimated by varying the hard-process renormalization and factorization scales, and by using the NNPDF replicas to estimate PDF uncertainties. For the \ttbar background process, additional modeling uncertainties are estimated by comparing the prediction from \POWHEG and \MGNLO~\cite{Alwall:2014hca}, by varying the renormalization scale used to simulate the initial- and final-state radiation, and by comparing the prediction using \PYTHIA and \HERWIG~\cite{Bellm:2015jjp} as parton showers. For the $Z+$jets background process, additional modeling uncertainties in the fraction of heavy-flavored associated jets are estimated by varying the resummation (QSF) and merging (CKKM) scales in \SHERPA~\cite{Bothmann:2019yzt,Schumann:2007mg}. For simplicity, the variations are estimated without detector simulation, but in a fiducial region identical to the one used in the search. In addition to generator-based estimates, uncertainties associated with the reweighting procedure in the $\ttbar$ and $Z$ regions are propagated to the final result.

Several uncorrelated sources of systematic uncertainties are considered for background events with non-prompt leptons. Statistical uncertainties coming from the limited number of events in the regions used to measure the prompt and non-prompt lepton efficiencies are generally negligible. Statistical uncertainties coming from the limited number of data events with at least one loose lepton are propagated to the final result. In addition to uncertainties of a statistical source, the following uncertainties are considered for background events with non-prompt leptons. Each uncertainty is estimated separately for electrons, muons, and hadronic taus.

\begin{itemize}
\item Prompt-lepton efficiency: the uncertainty is estimated by comparing the efficiency determined from simulated $Z+$jets events with the one estimated from simulated $\ttbar$ events.
\item Non-prompt-lepton composition: for electrons and muons, this uncertainty is estimated by comparing the non-prompt-lepton efficiency measured in the different analysis categories with the efficiency measured in a region with zero $b$- and $B$-jets. For hadronic taus, this uncertainty is obtained by using the same region without heavy-flavored jets to estimate variations on $r_{\text{QCD}}$ and $r_{\text{LJVT}}$.
\item Prompt-lepton subtraction: the uncertainty is estimated by varying the cross-section of processes with prompt leptons by 30\% in the regions used to estimate non-prompt efficiencies.
\end{itemize}

The uncertainty in the combined 2015-2018 integrated luminosity is 0.83\%~\cite{DAPR-2021-01}, obtained using the LUCID-2 detector for the main luminosity measurement, and complemented by measurements using the ID and the calorimeters. The uncertainty in the pileup model is determined by varying the estimate of the inelastic $pp$ cross-section between the value estimated by using only the LUCID-2 detector and the value estimated using only the ID.

\begin{table}
\centering
\caption{Impact of different group of systematic uncertainties on the upper limit on $(\sigma(H)/\sigma_{\text{SM}}(H)) \mathcal{B}(\haabbtautau)$. The different systematic uncertainties group are considered individually in each line of the table. The larger the difference relative  to the expected limit without systematic uncertainties (stat-only), the more important the uncertainty group is for the final result. The entries with ``--'' change by less than 1\% relative to the stat-only limit.}
\label{tab:syst_size}
\begin{tabular}{lccc} \toprule
\multirow{2}{*}{Uncertainty source} & \multicolumn{3}{c}{Expected limit on $(\sigma(H)/\sigma_{\text{SM}}(H)) \mathcal{B}(\haabbtautau)$} \\ \cline{2-4}
&~$m_a=12$~\GeV &~$m_a=25$~\GeV &~$m_a=60$~\GeV\\ \toprule
\textbf{Stat-only limit} & 1.34 & 1.79 & 3.00 \\ \midrule
\textbf{Observed limit} & 2.89 & 2.02 & 3.37 \\ \midrule
\hspace{0.5em}\textbf{MC statistics} & 1.42 & 1.81 & 3.04 \\ \midrule
\hspace{0.5em}\textbf{Experimental} & 2.72 & 1.94 & 3.21 \\ \midrule
\hspace{1em}\textbf{Detector response} & 2.43 & 1.84 & 3.03 \\
\hspace{1.5em}Luminosity and pileup & 1.37 & -- & -- \\
\hspace{1.5em}$b$-tagging & -- & 1.81 & --\\
\hspace{1.5em}$B$-tagging & 2.35 & -- & --\\
\hspace{1.5em}Jet and \met & -- & 1.83 & --\\
\hspace{1.5em}Electrons & 1.36 & -- & --\\
\hspace{1.5em}Muons & 1.35 & -- & --\\
\hspace{1.5em}Taus & -- & -- & --\\ \midrule
\hspace{1em}\textbf{Data-driven normalization} & 1.58 & 1.94 & 3.19\\
\hspace{1.5em}Non-prompt leptons & 1.58 & 1.85 & 3.16 \\
\hspace{1.5em}Non-prompt taus & -- & 1.86 & 3.10\\
\hspace{1.5em}MC reweighting & -- & -- & --\\ \midrule
\hspace{0.5em}\textbf{Theoretical modeling} & 1.38 & 1.89 & 3.04 \\
\hspace{1em}Signal & -- & -- & -- \\
\hspace{1em}Background & 1.37 & 1.87 & 3.03 \\ \bottomrule
\end{tabular}
\end{table}

\FloatBarrier
\section{Results}

Different $a$-boson mass hypotheses are tested $m_a=(12,16,20,25,30,40,50,60)\,\GeV$. Figures~\ref{fig:emu_1b_pnn_spectrum} and \ref{fig:emu_1B_pnn_spectrum} show the pNN distribution for each hypothesis in the three pNN bins used this search for the ($\emu$, 1b) and ($e\tauh)$ categories, respectively. While the $m^{\text{MMC}}(\tau\tau)$ window reduces the correlation between different hypotheses, the events in the histogram for different masses are not completely independent. No significant excess over the background expectation is observed for any $m_a$ hypothesis.  The largest deviation is a local $p$-value of 0.055 at $m_a$=20~\GeV.

Upper limits on $(\sigma(H)/\sigma_{\text{SM}}(H)) \mathcal{B}(\haabbtautau)$ are determined at 95\% confidence level (CL) using the CL$_{s}$ technique~\cite{Cowan:2010js} and the $\tilde{q}_{\mu}$ test statistic described in Section~\ref{sec:analysis_strategy}. The SM Higgs boson production cross-section used is $\sigma_{\text{SM}}(H) = 55.7\,\text{pb}$ \cite{LHCHXSWG3}. Table~\ref{tab:syst_size} shows the impact of dominant sources of systematic uncertainties in the expected upper limit for different $m_a$ hypotheses. The impact is estimated by the difference between the expected upper limit when only a group of uncertainties are considered and the case where no systematic uncertainty is considered. $B$-jet identification efficiency uncertainty is an important source of systematic uncertainty at low value of $m_a$ due to the difficulty in measuring the identification efficiency of merged $B$-jets with low mass and low $\pt$.  \Fig{\ref{fig:expected_limits_a}} compares the observed upper limits with the expected limits assuming the median value of $\tilde{q}_{\mu}$ under the background-only hypothesis.
\Fig{\ref{fig:expected_limits_b}} compares the expected limits for categories with heavy-flavor objects ($B$- or $b$-jets). The category using a $B$-jet is especially sensitive in the low-mass regime, where the $b\bar{b}$ system is collimated into a single object. Regarding the $\tau$-lepton-decay mode, the $\mu\tauh$ channel dominates the sensitivity, except for the low-mass regime where $e\mu$ dominates due to the isolation criterion, which is corrected for the presence of nearby leptons.

This is the first search in ATLAS for exotic decays of the Higgs boson in the final state $H \rightarrow aa \rightarrow b\bar{b}\tau^+\tau^-$. The observed limit is up to a factor of two better at low masses when compared to previous results~\cite{CMS-HIG-22-007} due to the use of a novel identification algorithm for low-mass merged $B$-jets. In several BSM models, final state with $b$-quarks have large branching ratio and the result provided here can place strict limits on the existence of exotic decays of the Higgs boson.

\begin{figure}[!tbp]
\centering
\subfloat[]{\includegraphics[width=\linewidth]{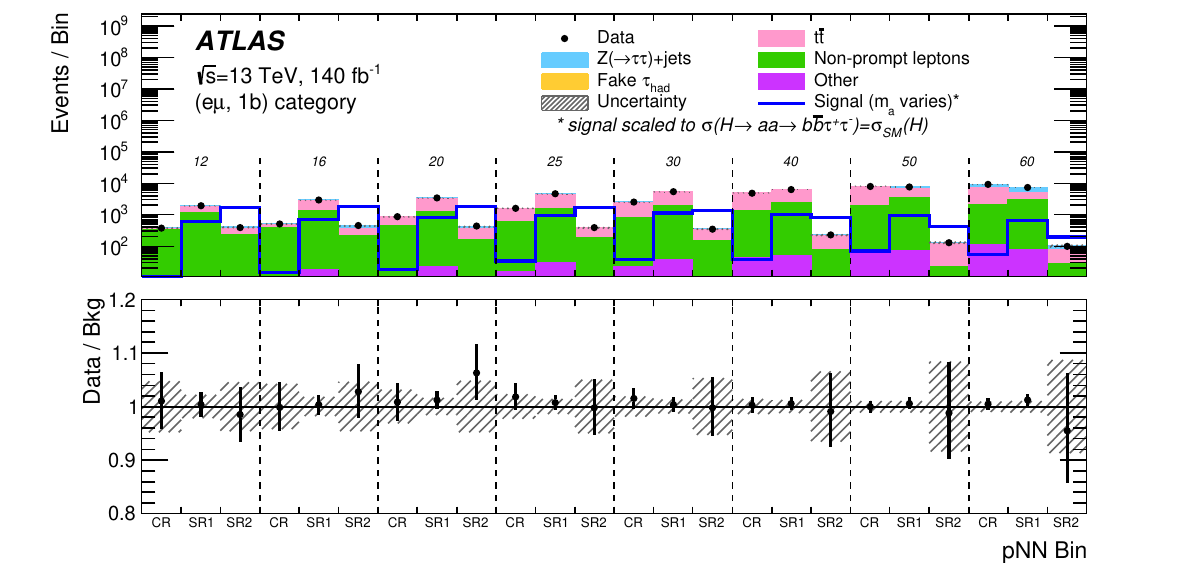}}\\
\subfloat[]{\includegraphics[width=\linewidth]{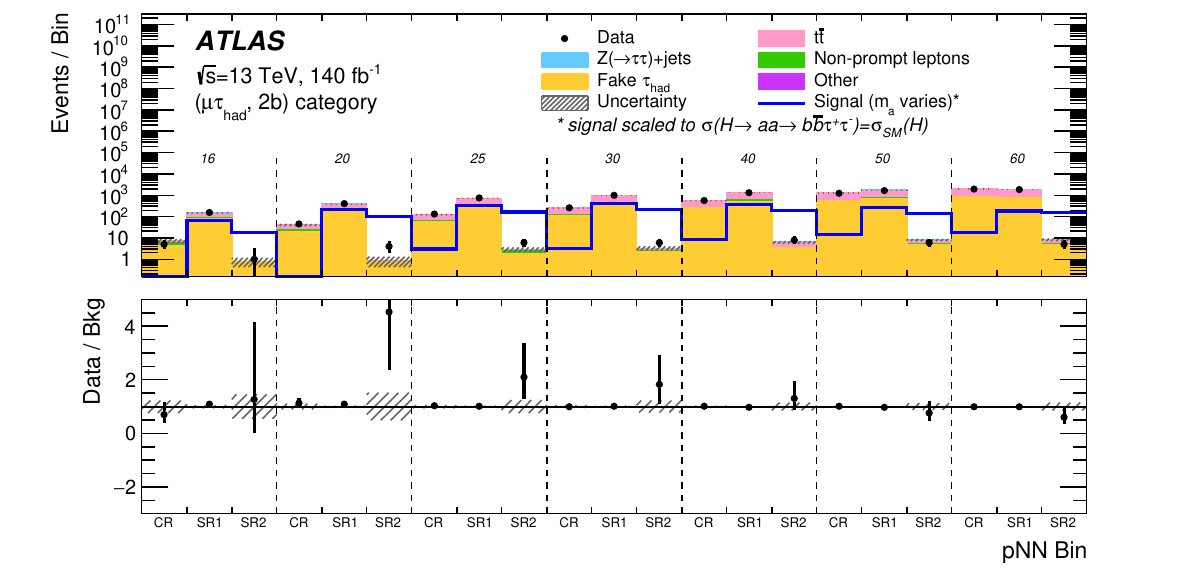}}
\caption{The pNN spectrum in the three bins are shown separately for each mass hypothesis in the (a) (\emu, $1b$) and (b) (\muhad, $2b$) category.  The signal shape (normalized to $\mathcal{B}(\haabbtautau)=1$) for each corresponding mass hypothesis is overlaid on top of the SM prediction. The bins divided by vertical dashed lines are used for testing different $m_a$ hypotheses indicated as a text in the figure. Bins used for different $m_a$ hypotheses are not statistically independent. Note that the 12~\GeV mass point is not shown for the (\muhad, $2b$) category because the signal acceptance is very low. The hashed area represents the total uncertainty of the background.}
\label{fig:emu_1b_pnn_spectrum}
\end{figure}

\begin{figure}[!tbp]
\centering
\subfloat[]{\includegraphics[width=\linewidth]{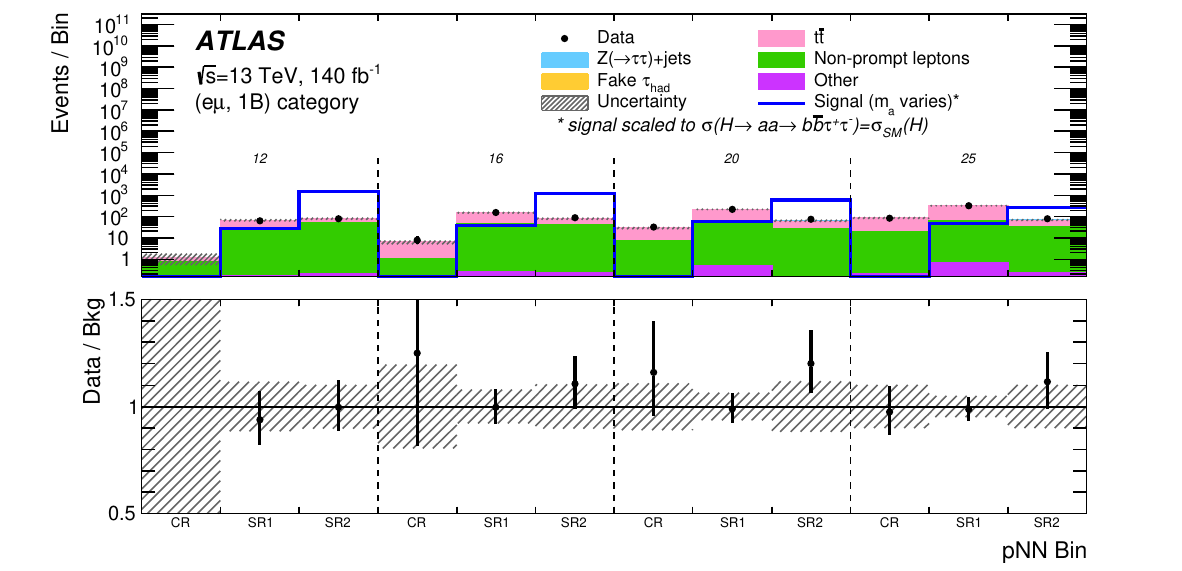}}\\
\subfloat[]{\includegraphics[width=\linewidth]{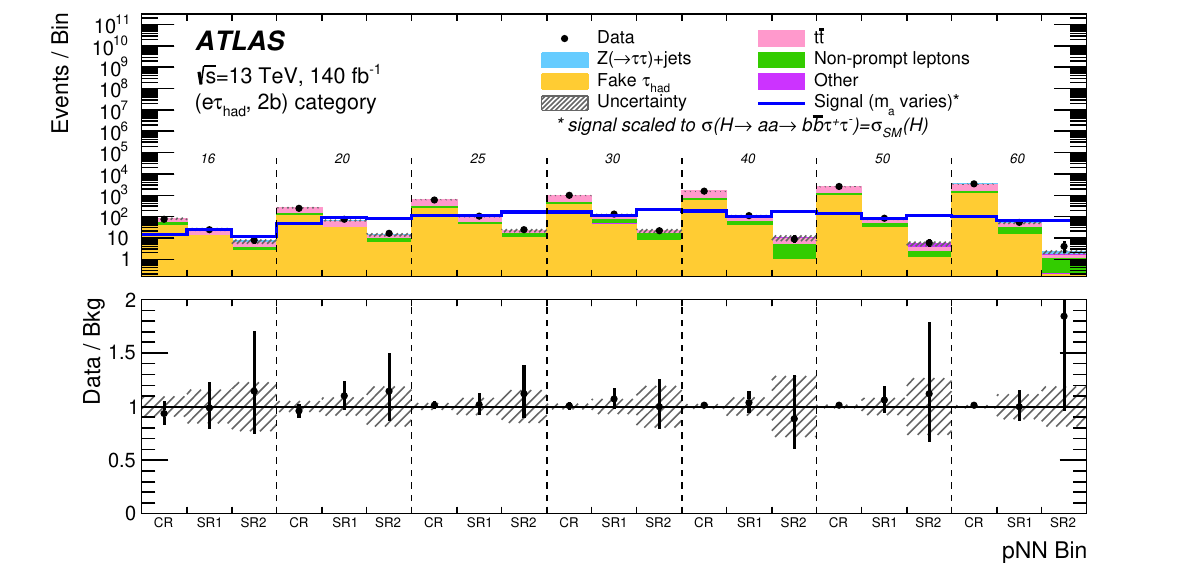}}
\caption{The pNN spectrum in the three bins are shown separately for each mass hypothesis in the (a) (\emu, $1B$) and (b) (\ehad, $2b$) categories.  The signal shape (normalized to $\mathcal{B}(H\rightarrow aa\rightarrow bb\tau\tau)=1)$ for each corresponding mass hypothesis is overlaid on top of the SM prediction. The bins divided by vertical dashed lines are used for testing different $m_a$ hypotheses indicated as a text in the figure. Bins used for different $m_a$ hypotheses are not statistically independent. The hashed area represents the total uncertainty of the background.}
\label{fig:emu_1B_pnn_spectrum}
\end{figure}

\begin{figure}[!tbp]
\centering
\subfloat[]{\label{fig:expected_limits_a}\includegraphics[height=5.8cm]{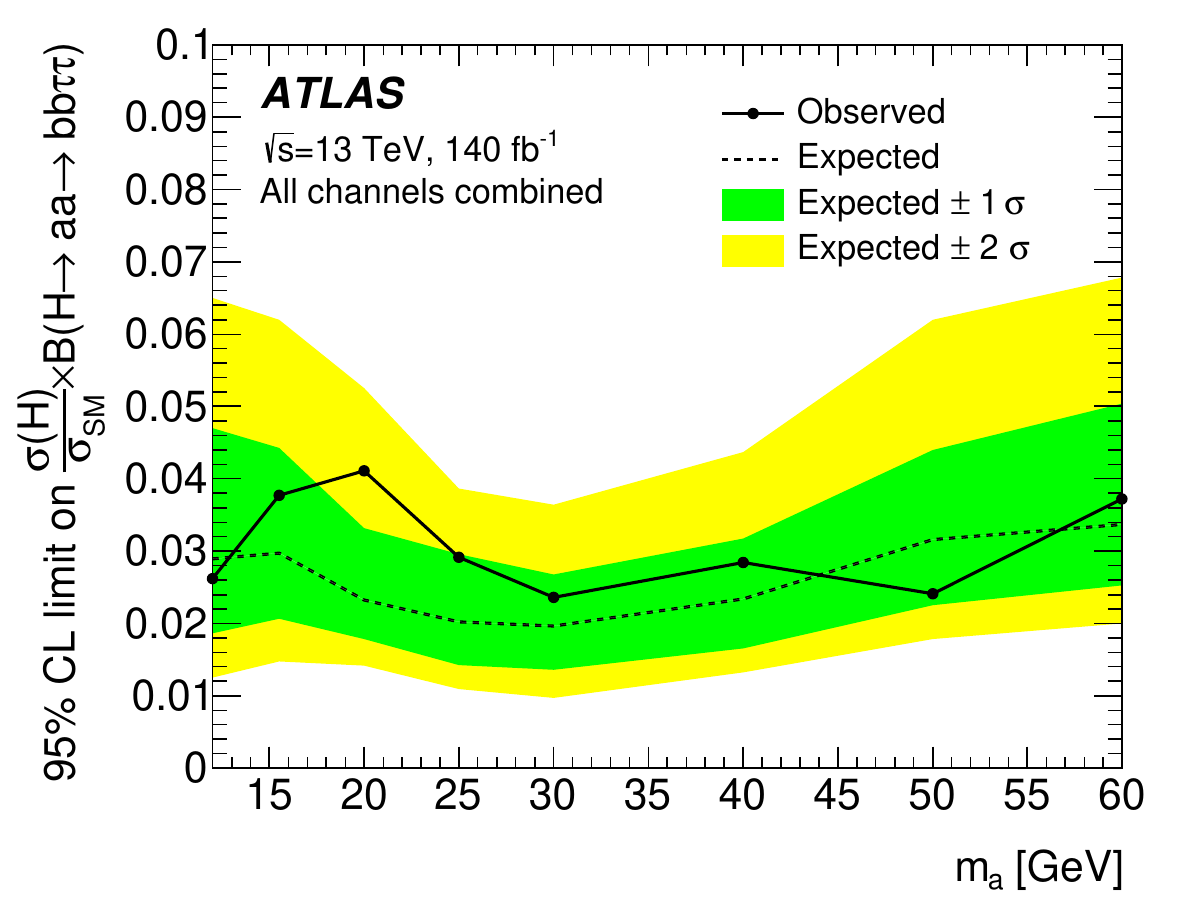}}%
\subfloat[]{\label{fig:expected_limits_b}\includegraphics[height=5.8cm]{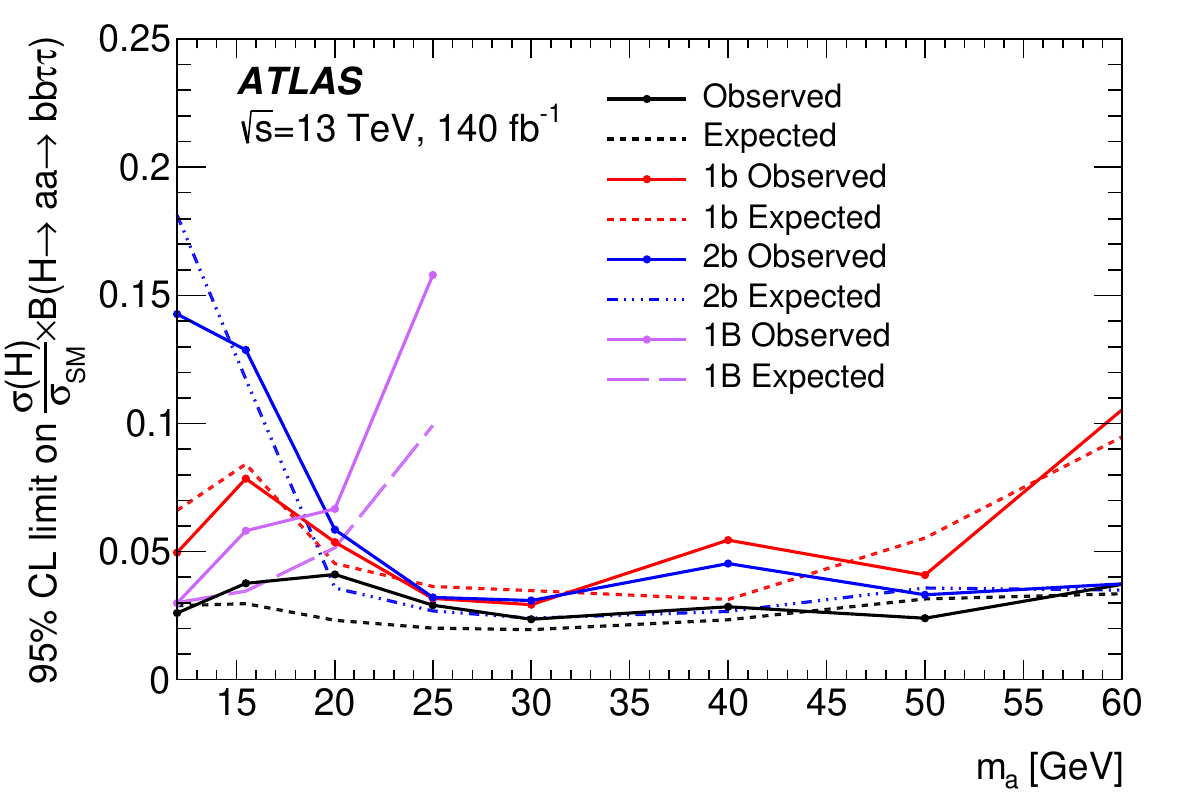}}
\caption{The observed (solid) 95\% C.L. upper limits on $(\sigma(H)/\sigma_{\text{SM}}(H)) \mathcal{B}(\haabbtautau)$ as a function of $m_a$ and the expected (dashed) limits under the background-only  hypothesis when (a) combining all categories and (b) considering different categories based on the heavy-flavor objects separately. In the combined plot (a) the inner green and outer yellow shaded bands show the $\pm 1\sigma$ and $\pm 2\sigma$ uncertainties of the expected limits. The mass hypothesis $m_a$ is probed between 12 and 60~\GeV for the values shown with markers. A linear interpolation validated with MC simulation between adjacent mass points is used.
}%
\end{figure}

\FloatBarrier
\section{Conclusion}

This paper presents the first search by the ATLAS experiment for exotic decays of the Higgs boson into a pair of pseudoscalar particles, $H\rightarrow aa$,
where the $a$-boson decays into two $b$-quarks and two $\tau$-leptons.
The analysis is performed using the 140 \ifb\ of proton--proton collision data at a center-of-mass energy of $13$ \TeV recorded by the ATLAS detector at the LHC between 2015 and 2018.
The search makes use of heavy-flavor-tagging techniques to target collimated $a\rightarrow b\bar{b}$ decays and a mass-parameterized neural network to enhance the sensitivity. The analysis categorizes events depending on the $\tau$-lepton decay modes: $e\mu$, $e\tauh$, and $\mu\tauh$, as well as the strategy for identifying heavy-flavor jets: a merged \bjet pair, $B$-jet, and one or two single $b$-jets.
No significant excess above the SM background expectation is observed, and upper limits at 95\% confidence level are set on $\mathcal{B}(\haabbtautau)$ of $2.2\% - 3.9\%$ for pseudoscalar mass values in the range $12~\GeV \leq m_{a} \leq 60~\GeV$, assuming the SM Higgs boson production cross-section. These results contribute to the broad program of searches for $H\rightarrow aa$ decays in ATLAS and can be used to set constraints on a variety of BSM scenarios featuring such exotic Higgs decays.

\section*{Acknowledgements}
%

%
%

%
%

We thank CERN for the very successful operation of the LHC and its injectors, as well as the support staff at
CERN and at our institutions worldwide without whom ATLAS could not be operated efficiently.

The crucial computing support from all WLCG partners is acknowledged gratefully, in particular from CERN, the ATLAS Tier-1 facilities at TRIUMF/SFU (Canada), NDGF (Denmark, Norway, Sweden), CC-IN2P3 (France), KIT/GridKA (Germany), INFN-CNAF (Italy), NL-T1 (Netherlands), PIC (Spain), RAL (UK) and BNL (USA), the Tier-2 facilities worldwide and large non-WLCG resource providers. Major contributors of computing resources are listed in Ref.~\cite{ATL-SOFT-PUB-2023-001}.

We gratefully acknowledge the support of ANPCyT, Argentina; YerPhI, Armenia; ARC, Australia; BMWFW and FWF, Austria; ANAS, Azerbaijan; CNPq and FAPESP, Brazil; NSERC, NRC and CFI, Canada; CERN; ANID, Chile; CAS, MOST and NSFC, China; Minciencias, Colombia; MEYS CR, Czech Republic; DNRF and DNSRC, Denmark; IN2P3-CNRS and CEA-DRF/IRFU, France; SRNSFG, Georgia; BMBF, HGF and MPG, Germany; GSRI, Greece; RGC and Hong Kong SAR, China; ISF and Benoziyo Center, Israel; INFN, Italy; MEXT and JSPS, Japan; CNRST, Morocco; NWO, Netherlands; RCN, Norway; MNiSW, Poland; FCT, Portugal; MNE/IFA, Romania; MESTD, Serbia; MSSR, Slovakia; ARRS and MIZ\v{S}, Slovenia; DSI/NRF, South Africa; MICINN, Spain; SRC and Wallenberg Foundation, Sweden; SERI, SNSF and Cantons of Bern and Geneva, Switzerland; MOST, Taipei; TENMAK, T\"urkiye; STFC, United Kingdom; DOE and NSF, United States of America.

Individual groups and members have received support from BCKDF, CANARIE, CRC and DRAC, Canada; CERN-CZ, FORTE and PRIMUS, Czech Republic; COST, ERC, ERDF, Horizon 2020, ICSC-NextGenerationEU and Marie Sk{\l}odowska-Curie Actions, European Union; Investissements d'Avenir Labex, Investissements d'Avenir Idex and ANR, France; DFG and AvH Foundation, Germany; Herakleitos, Thales and Aristeia programmes co-financed by EU-ESF and the Greek NSRF, Greece; BSF-NSF and MINERVA, Israel; NCN and NAWA, Poland; La Caixa Banking Foundation, CERCA Programme Generalitat de Catalunya and PROMETEO and GenT Programmes Generalitat Valenciana, Spain; G\"{o}ran Gustafssons Stiftelse, Sweden; The Royal Society and Leverhulme Trust, United Kingdom.

In addition, individual members wish to acknowledge support from Armenia: Yerevan Physics Institute (FAPERJ); CERN: European Organization for Nuclear Research (CERN PJAS); Chile: Agencia Nacional de Investigaci\'on y Desarrollo (FONDECYT 1230812, FONDECYT 1230987, FONDECYT 1240864); China: Chinese Ministry of Science and Technology (MOST-2023YFA1605700), National Natural Science Foundation of China (NSFC - 12175119, NSFC 12275265, NSFC-12075060); Czech Republic: Czech Science Foundation (GACR - 24-11373S), Ministry of Education Youth and Sports (FORTE CZ.02.01.01/00/22\_008/0004632), PRIMUS Research Programme (PRIMUS/21/SCI/017); EU: H2020 European Research Council (ERC - 101002463); European Union: European Research Council (ERC - 948254, ERC 101089007), Horizon 2020 Framework Programme (MUCCA - CHIST-ERA-19-XAI-00), European Union, Future Artificial Intelligence Research (FAIR-NextGenerationEU PE00000013), Italian Center for High Performance Computing, Big Data and Quantum Computing (ICSC, NextGenerationEU); France: Agence Nationale de la Recherche (ANR-20-CE31-0013, ANR-21-CE31-0013, ANR-21-CE31-0022), Investissements d'Avenir Labex (ANR-11-LABX-0012); Germany: Baden-Württemberg Stiftung (BW Stiftung-Postdoc Eliteprogramme), Deutsche Forschungsgemeinschaft (DFG - 469666862, DFG - CR 312/5-2); Italy: Istituto Nazionale di Fisica Nucleare (ICSC, NextGenerationEU); Japan: Japan Society for the Promotion of Science (JSPS KAKENHI JP22H01227, JSPS KAKENHI JP22H04944, JSPS KAKENHI JP22KK0227, JSPS KAKENHI JP23KK0245); Netherlands: Netherlands Organisation for Scientific Research (NWO Veni 2020 - VI.Veni.202.179); Norway: Research Council of Norway (RCN-314472); Poland: Polish National Agency for Academic Exchange (PPN/PPO/2020/1/00002/U/00001), Polish National Science Centre (NCN 2021/42/E/ST2/00350, NCN OPUS nr 2022/47/B/ST2/03059, NCN UMO-2019/34/E/ST2/00393, UMO-2020/37/B/ST2/01043, UMO-2021/40/C/ST2/00187, UMO-2022/47/O/ST2/00148, UMO-2023/49/B/ST2/04085); Slovenia: Slovenian Research Agency (ARIS grant J1-3010); Spain: Generalitat Valenciana (Artemisa, FEDER, IDIFEDER/2018/048), Ministry of Science and Innovation (MCIN \& NextGenEU PCI2022-135018-2, MICIN \& FEDER PID2021-125273NB, RYC2019-028510-I, RYC2020-030254-I, RYC2021-031273-I, RYC2022-038164-I), PROMETEO and GenT Programmes Generalitat Valenciana (CIDEGENT/2019/027); Sweden: Swedish Research Council (Swedish Research Council 2023-04654, VR 2018-00482, VR 2022-03845, VR 2022-04683, VR 2023-03403, VR grant 2021-03651), Knut and Alice Wallenberg Foundation (KAW 2018.0157, KAW 2018.0458, KAW 2019.0447, KAW 2022.0358); Switzerland: Swiss National Science Foundation (SNSF - PCEFP2\_194658); United Kingdom: Leverhulme Trust (Leverhulme Trust RPG-2020-004), Royal Society (NIF-R1-231091); United States of America: U.S. Department of Energy (ECA DE-AC02-76SF00515), Neubauer Family Foundation.

%
%


%
\clearpage
\printbibliography
\clearpage
 
\begin{flushleft}
\hypersetup{urlcolor=black}
{\Large The ATLAS Collaboration}

\bigskip

\AtlasOrcid[0000-0002-6665-4934]{G.~Aad}$^\textrm{\scriptsize 104}$,
\AtlasOrcid[0000-0001-7616-1554]{E.~Aakvaag}$^\textrm{\scriptsize 17}$,
\AtlasOrcid[0000-0002-5888-2734]{B.~Abbott}$^\textrm{\scriptsize 123}$,
\AtlasOrcid[0000-0002-0287-5869]{S.~Abdelhameed}$^\textrm{\scriptsize 119a}$,
\AtlasOrcid[0000-0002-1002-1652]{K.~Abeling}$^\textrm{\scriptsize 56}$,
\AtlasOrcid[0000-0001-5763-2760]{N.J.~Abicht}$^\textrm{\scriptsize 50}$,
\AtlasOrcid[0000-0002-8496-9294]{S.H.~Abidi}$^\textrm{\scriptsize 30}$,
\AtlasOrcid[0009-0003-6578-220X]{M.~Aboelela}$^\textrm{\scriptsize 45}$,
\AtlasOrcid[0000-0002-9987-2292]{A.~Aboulhorma}$^\textrm{\scriptsize 36e}$,
\AtlasOrcid[0000-0001-5329-6640]{H.~Abramowicz}$^\textrm{\scriptsize 154}$
% ...
\bigskip
\\

% ...
$^{17}$Department for Physics and Technology, University of Bergen, Bergen; Norway.\\
% ...
$^{30}$Physics Department, Brookhaven National Laboratory, Upton NY; United States of America.\\
% ...
$^{36}$$^{(a)}$Facult\'e des Sciences Ain Chock, R\'eseau Universitaire de Physique des Hautes Energies - Universit\'e Hassan II, Casablanca;$^{(b)}$Facult\'{e} des Sciences, Universit\'{e} Ibn-Tofail, K\'{e}nitra;$^{(c)}$Facult\'e des Sciences Semlalia, Universit\'e Cadi Ayyad, LPHEA-Marrakech;$^{(d)}$LPMR, Facult\'e des Sciences, Universit\'e Mohamed Premier, Oujda;$^{(e)}$Facult\'e des sciences, Universit\'e Mohammed V, Rabat;$^{(f)}$Institute of Applied Physics, Mohammed VI Polytechnic University, Ben Guerir; Morocco.\\
% ...
$^{45}$Physics Department, Southern Methodist University, Dallas TX; United States of America.\\
% ...
$^{50}$Fakult\"{a}t Physik , Technische Universit{\"a}t Dortmund, Dortmund; Germany.\\
% ...
$^{56}$II. Physikalisches Institut, Georg-August-Universit\"{a}t G\"ottingen, G\"ottingen; Germany.\\
% ...
$^{104}$CPPM, Aix-Marseille Universit\'e, CNRS/IN2P3, Marseille; France.\\
% ...
$^{119}$$^{(a)}$New York University Abu Dhabi, Abu Dhabi;$^{(b)}$United Arab Emirates University, Al Ain; United Arab Emirates.\\
% ...
$^{123}$Homer L. Dodge Department of Physics and Astronomy, University of Oklahoma, Norman OK; United States of America.\\
% ...
$^{154}$Raymond and Beverly Sackler School of Physics and Astronomy, Tel Aviv University, Tel Aviv; Israel.\\
% ...
$^{a}$ Also Affiliated with an institute covered by a cooperation agreement with CERN.\\
$^{b}$ Also at An-Najah National University, Nablus; Palestine.\\
$^{c}$ Also at Borough of Manhattan Community College, City University of New York, New York NY; United States of America.\\
$^{d}$ Also at Center for Interdisciplinary Research and Innovation (CIRI-AUTH), Thessaloniki; Greece.\\
$^{e}$ Also at Centro Studi e Ricerche Enrico Fermi; Italy.\\
$^{f}$ Also at CERN, Geneva; Switzerland.\\
% ...

\end{flushleft}


%

%
%

%

\end{document}